\documentclass[%
 reprint,
 superscriptaddress,
 amsmath,amssymb,
 aps,
 floatfix
]{revtex4-2}

\usepackage{graphicx}
\usepackage{dcolumn}
\usepackage{bm}
\usepackage{physics}
\usepackage{csquotes}
\usepackage{mathtools}
\usepackage{amsfonts}
\usepackage{amssymb}
\usepackage{amsmath}
\usepackage[caption=false,labelformat=empty]{subfig}
\usepackage{diagbox}
\usepackage{dsfont}
\usepackage{hyperref}
\usepackage{placeins}
\usepackage{xspace}
\hypersetup{linktocpage,colorlinks,citecolor={blue},pdfdisplaydoctitle=true,pdfpagemode=UseOutlines,bookmarksnumbered=true}
\usepackage{verbatim}

\newcommand*{\maj}[2]{\ensuremath{#1^{(#2)}}}
\newcommand*{\dgr}{\ensuremath{^{\dagger}}}

\newcommand*{\gauge}{\ensuremath{\mathcal{Q}}}

\newcommand*{\gammain}{\ensuremath{\Gamma_{\text{in}}}}
\newcommand*{\gammainmod}{\ensuremath{\widetilde{\Gamma}_{\text{in}}}}

\newcommand*{\Z}[1]{\ensuremath{\mathds{Z}_{#1}}}
\newcommand*{\normsq}{\ensuremath{\abs{\Psi(\gauge)}^2}}

\newcommand*{\id}{\ensuremath{\mathds{1}}}
\newcommand*{\xarg}{\ensuremath{\vb{x}}}

\newcommand*{\tran}{\ensuremath{^T}}

\newcommand*{\inv}{\ensuremath{^{-1}}}

\DeclareMathOperator{\Pf}{Pf}

\begin{document}

\title{Finding the ground state of a lattice gauge theory with fermionic tensor networks: a $2+1d$ \texorpdfstring{\Z{2}}{Z(2)} demonstration}

\author{Patrick Emonts}
\affiliation{Max-Planck Institute of Quantum Optics, Hans-Kopfermann-Str. 1, 85748 Garching, Germany}
\affiliation{Munich Center for Quantum Science and Technology (MCQST), Schellingstr. 4, D-80799 München}
\affiliation{Instituut-Lorentz, Universiteit Leiden, P.O. Box 9506, 2300 RA Leiden, The Netherlands}
\author{Ariel Kelman}
\affiliation{Racah Institute of Physics, The Hebrew University of Jerusalem, Givat Ram, Jerusalem 91904, Israel}
\author{Umberto Borla}
\affiliation{Munich Center for Quantum Science and Technology (MCQST), Schellingstr. 4, D-80799 München}
\affiliation{Physik-Department, Technische Universit\"{a}t M\"{u}nchen, 85748 Garching, Germany}
\author{Sergej Moroz}
\affiliation{Department of Engineering and Physics, Karlstad University, Karlstad, Sweden}
\author{Snir Gazit}
\affiliation{Racah Institute of Physics, The Hebrew University of Jerusalem, Givat Ram, Jerusalem 91904, Israel}
 \affiliation{The Fritz Haber Research Center for Molecular Dynamics, The Hebrew University of Jerusalem, Jerusalem 91904, Israel}
\author{Erez Zohar}
\affiliation{Racah Institute of Physics, The Hebrew University of Jerusalem, Givat Ram, Jerusalem 91904, Israel}

\date{\today}

\begin{abstract}
Tensor network states, and in particular Projected Entangled Pair States (PEPS) have been a strong ansatz for the variational study of complicated quantum many-body systems, thanks to their built-in entanglement entropy area law. 
In this work, we use a special kind of PEPS - Gauged Gaussian Fermionic PEPS (GGFPEPS) - to find the ground state of $2+1$ dimensional pure $\mathbb{Z}_2$ lattice gauge theories for a wide range of coupling constants. 
We do so by combining PEPS methods with Monte-Carlo computations, allowing for efficient contraction of the PEPS and computation of correlation functions. 
Previously, such numerical computations involved the calculation of the Pfaffian of a matrix scaling with the system size, forming a severe bottleneck; in this work we show how to overcome this problem. This paves the way for applying the method we propose and benchmark here to other gauge groups, higher dimensions, and models with fermionic matter, in an efficient, sign-problem-free way.
\end{abstract}
\maketitle

\section{Introduction\label{sec:introduction}}
Tensor network states (TNS), and matrix product states (MPS) in particular, have been very fruitful in dealing with complicated quantum many body models in condensed matter physics. 
MPS are ansatz states with a built-in area law for the entanglement entropy~\cite{orus_practical_2014,cirac_renormalization_2009,cirac_matrix_2021} equipped with algorithms which scale polynomially with the system size, rather than exponentially. 
This opens the way for a very efficient ground state search~\cite{white_density_1992,schollwock_density-matrix_2011}, as well as studying the dynamics~\cite{daley_time-dependent_2004} and thermal states \cite{verstraete_matrix_2004,zwolak_mixed-state_2004} of quantum many-body systems. 
The idea of tensor networks can be extended to higher spatial dimensions, when MPS are generalized to projected entangled pair states (PEPS)~\cite{jordan_classical_2008,cirac_renormalization_2009,corboz_simulation_2010,cirac_matrix_2021}. 
However, in spite of the great numerical success of MPS, PEPS in higher dimensions are generally very difficult to handle (i.e. contract) numerically; while they have been successfully applied to two dimensional models in some cases (see, e.g.~\cite{corboz_competing_2014,niesen_tensor_2017}), the computational time required for contracting PEPS generally scales unfavorably when increasing the spatial dimension to more than one \cite{verstraete_renormalization_2004}.

Gauge theories are at the heart of our modern physics.
They play a central role in the standard model of particle physics, giving a local description of the fundamental interactions~\cite{peskin_introduction_1999}. 
In condensed matter physics, they offer effective and emergent descriptions of intriguing many-body phenomena~\cite{fradkin_field_2013}. 
The local symmetries upon which they are based allow for a simple and local formulation of complicated interactions, and give rise to a highly constrained Hilbert space. 
Gauge theories exhibit some fascinating features, such as running coupling, which can be both useful and challenging: for example, in Quantum Chromodynamics (QCD), the theory of the strong nuclear force, the high energy limit is asymptotically free~\cite{gross_asymptotically_1973}, allowing one to study collider physics using perturbation theory. 
The low-energy side, on the other hand, is strongly interacting and highly nonperturbative, giving rise to beautiful yet challenging physical phenomena such as quark confinement~\cite{wilson_confinement_1974,greensite_confinement_2003} involving many open puzzles.

A very successful method to address nonperturbative gauge theory physics has been within the framework of lattice gauge theories~\cite{wilson_confinement_1974,kogut_hamiltonian_1975,polyakov_quark_1977,kogut_introduction_1979}. 
Based on Monte-Carlo simulations of Wick-rotated path integrals in Euclidean spacetime~\cite{creutz_monte_1980,creutz_monte_1983}, many challenging computations of static quantities in gauge theories have been carried out (for example, much of our knowledge of the hadronic spectrum these days is the result of lattice Monte-Carlo computations~\cite{aoki_flag_2021}).
On the other hand, these methods do not allow us to directly study real-time evolution, and thus are not suitable for dynamical analysis. Additionally, in many important physical scenarios with a finite density of fermionic matter, the sign problem~\cite{troyer_computational_2005} makes the use of Monte-Carlo methods impossible.

In the last decade, some quantum information based methods have been proposed to deal with these issues. 
First, quantum simulation~\cite{feynman_simulating_1982} offers one approach to simulate this challenging area of physics using atomic, molecular, optical and solid state devices; one can design~\cite{wiese_ultracold_2013,zohar_quantum_2016,dalmonte_lattice_2016,banuls_simulating_2020,aidelsburger_cold_2022,bass_quantum_2022,klco_standard_2022}
and build~\cite{martinez_real-time_2016,kokail_self-verifying_2019,schweizer_floquet_2019,mil_scalable_2020,yang_observation_2020,semeghini_probing_2021,zhou_thermalization_2022,riechert_engineering_2022}
quantum simulators of lattice gauge theories which are free of the above-mentioned issues, and this is a very promising and exciting avenue of research. Tensor network states, on which this work focuses, provide another quantum-information-based way to deal with them~\cite{banuls_review_2020,montangero_loop-free_2022}.

In a single space dimension, Matrix product states, combined with DMRG (density matrix renormalization group) techniques~\cite{white_density_1992,schollwock_density-matrix_2005}
have been successfully applied to lattice gauge theories, both in particle  and condensed matter physics (see reviews \cite{banuls_review_2020,dalmonte_lattice_2016,banuls_simulating_2020,lumia_two-dimensional_2022} and references therein). 
A reliable but limited approach is to extend MPS-based algorithms to two-dimensional systems defined on ladders or cylinders. 
While this approach has been  successfully applied to study gapped phases of matter and the transitions between them (e.g.,  \cite{tschirsich_phase_2019,gonzalez-cuadra_robust_2020,borla_quantum_2022,brenig_spinless_2022} and others), the computational cost of this approach grows exponentially with additional dimensions, making the extrapolation to the thermodynamic limit generally impossible.
In higher dimensions, tensor network techniques have been used for studying lattice gauge theories.
Lattice gauge theories have been studied numerically using various methods in \cite{tagliacozzo_entanglement_2011,tagliacozzo_tensor_2014,crone_detecting_2020,robaina_simulating_2021,felser_two-dimensional_2020,montangero_loop-free_2022,magnifico_lattice_2021}. 

In parallel, some analytical work has been done out on the properties of gauge invariant PEPS, and in particular on gauging mechanisms~\cite{haegeman_gauging_2015,zohar_building_2016} which allow one to minimally couple gauge fields to matter-only PEPS in a straight-forward manner even for arbitrary groups and space dimensions. 
The latter method, when combined with Gaussian fermionic PEPS~\cite{kraus_fermionic_2010}, was used for introducing Gauged Gaussian Fermionic PEPS (GGFPEPS)~\cite{zohar_fermionic_2015,zohar_projected_2016}. 
In such states, a free (Gaussian) fermionic matter state with a global symmetry is gauged in a way analogous to conventional Hamiltonian-level minimal coupling techniques~\cite{emonts_gauss_2020}, resulting in a  physically relevant state which describes quantum matter and gauge fields interacting in a gauge-invariant way. 
Moreover, it was shown that this special class of PEPS allows one to contract the states and compute correlation functions efficiently using Monte-Carlo sampling~\cite{zohar_combining_2018}. 
For this method, the sampling probability is shown to always depend on the norm of a quantum state, and hence it is sign-problem-free. 
Therefore one can use GGFPEPS as ansatz states for variational Monte-Carlo~\cite{sorella_generalized_2001,sorella_wave_2005} ground state search  of lattice gauge theory Hamiltonians, overcoming the common sign problem  experienced by lattice gauge theories, as well as the difficulty of contracting PEPS in more than a single space dimension.
Previous works have variations of tensor network states as variational Ansatz for Monte Carlo, e.g. ~\cite{sandvik_variational_2007,schuch_simulation_2008,ferris_perfect_2012}.
Here, we construct anstatz states that are locally gauge invariant and adapted to the gauge group in question.

In previous work~\cite{emonts_variational_2020}, this method was used to find the ground state of a pure $\mathbb{Z}_3$ lattice gauge theory. 
While successful for most values of the coupling constant, one important issue with the algorithm was the need for computation of Pfaffians~\cite{bravyi_lagrangian_2005} of matrices whose dimensions scales with the size of the physical system. 
This scaling posed a serious bottleneck on using GGFPEPS as variational ansatz states~\cite{wimmer_algorithm_2012}. 
In this work we present a simple way to overcome this problem, and demonstrate its use for finding the ground state of a $\mathbb{Z}_2$ pure gauge theory~\cite{wegner_duality_1971,elitzur_phase_1979,horn_hamiltonian_1979,trebst_breakdown_2007,tupitsyn_topological_2010}. 
While this computational problem on its own does not suffer from the sign problem, the techniques demonstrated in this work can be generalized in a straight-forward way to cases which do suffer from it (e.g., coupling the very same theory to physical fermions with an odd number of flavors and imposing the gauge constraint \cite{gazit_emergent_2017,borla_quantum_2022}). 

This paper is structured as follows: 
In Secs.~\ref{sec:phys_system} and~\ref{sec:construction}, we introduce $\Z{N}$ gauge theories with a special focus on $\Z{2}$ and constructing the GGFPEPS ansatz state.
Subsequently, the GGFPEPS are minimized with the algorithm described in Sec.~\ref{sec:algorithm}.
Numerical results are presented in Sec.~\ref{sec:results} and we conclude in Sec.~\ref{sec:conclusions}.

\section{Physical System\label{sec:phys_system}}
We focus on pure gauge $\mathbb{Z}_N$ lattice gauge theories in the $2+1d$ Hamiltonian framework; i.e.  the physical system is a two dimensional spatial lattice, with gauge field degrees of freedom occupying its links. Due to the absence of dynamical matter, no physical degrees of freedom are associated with the sites.

Each link $\ell = (\mathbf{x},i)$ (labelled by the starting site $\mathbf{x}$ and a direction $i \in \{1,2$\}) hosts an $N$ dimensional Hilbert space. On each link $\ell$ we introduce two  operators, which satisfy the following conditions
\begin{equation}
  \begin{aligned}
    &P_\ell^N=Q_\ell^N=1,&& P_\ell\dgr P_\ell=Q_\ell\dgr Q_\ell=1\\
    &P_\ell Q_\ell P_\ell\dgr  =e^{i\delta }Q_\ell, && \delta=\frac{2\pi}{N} \, 
  \end{aligned}
  \label{eq:ZN_algebra}
\end{equation}
-- $N$th roots of unity, unitarity, and $\mathbb{Z}_N$ algebra respectively~\cite{horn_hamiltonian_1979}.
If we label the eigenstates of $P$ by $\left\{\left|p\right\rangle\right\}_{p=0}^{N-1}$, for which
\begin{equation}
P\left|p\right\rangle = e^{i p \delta}\left|p\right\rangle,
\end{equation}
we can immediately deduce that $Q$ is a unitary and periodic ($\left|N\right\rangle=\left|0\right\rangle$) raising operator,
\begin{equation}
Q\left|p\right\rangle = \left|p+1\right\rangle.
\end{equation}
Similarly, one can show that the eigenstates of $Q$, $\left\{\left|q\right\rangle\right\}_{q=0}^{N-1}$, for which
\begin{equation}
Q\left|q\right\rangle = e^{i q \delta}\left|p\right\rangle,
\end{equation}
are unitarily and periodically lowered by $P$,
\begin{equation}
P\left|q\right\rangle = \left|q-1\right\rangle.
\end{equation}

\begin{figure}
    \centering
    \includegraphics[width=0.5\columnwidth]{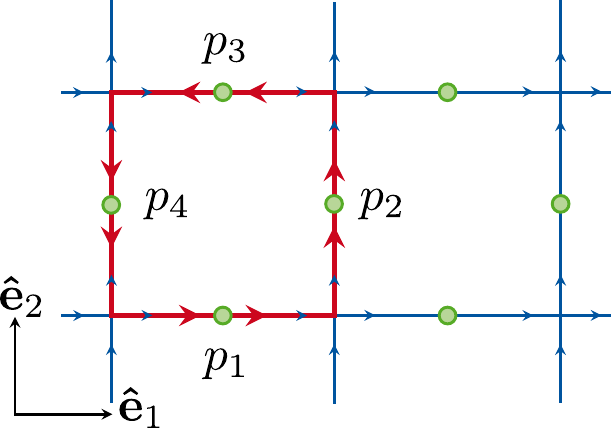}
    \caption{Illustration of the naming conventions on the lattice.
    Green dots illustrate the positions of the gauge fields.
    The small blue arrows indicate the direction of the discrete divergence at every vertex.
    The red square is an oriented plaquette (indicated by red arrows) with four links $p_1, p_2, p_3$ and $p_4$. 
    }
    \label{fig:lattice}
\end{figure}

The dynamics is given by the Hamiltonian~\cite{horn_hamiltonian_1979}
\begin{equation} 
    \begin{aligned}
        H &= \frac{\lambda}{2}\sum_\ell\left[2-(P_\ell+P_\ell\dgr)\right]
        \\& + \frac{1}{2\lambda}\sum_p\left[2-(Q_{p_1}\dgr Q_{p_2}\dgr Q_{p_3}Q_{p_4}+\text{H.c.})\right]
    \end{aligned} 
    \label{eq:hamilton_ZN}
\end{equation}
where the second sum is over plaquettes $p$ - unit squares of the lattice - and $p_1,p_2,p_3,p_4$ refer to the four links around $p$ (see Figure~\ref{fig:lattice} for an illustration of the notation).

\begin{figure}
    \centering
    \includegraphics[width=0.5\columnwidth]{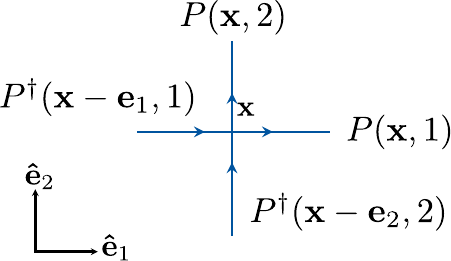}
    \caption{Illustration of the Gauss law operator.
    It is written in terms of a discrete divergence on the lattice.}
    \label{fig:gauss_law}
\end{figure}

The Hamiltonian is gauge invariant; that it, it is invariant under \emph{local} unitary transformations of the form
\begin{equation}
V\left(\xarg\right) =   P\left(\xarg,1\right)P\left(\xarg,2\right)P^{\dagger}\left(\xarg-\vu{e}_1,1\right)P^{\dagger}\left(\xarg-\vu{e}_2,2\right)
\end{equation}
(where $\vu{e}_i$ is a lattice vector in the $i \in \{1,2\}$ direction; see Figure \ref{fig:gauss_law}). 
Since $\left[H,V\left(\xarg\right)\right]=\left[V\left(\xarg\right),V\left(\vb{y}\right)\right]=0$ for any lattice sites $\xarg,\vb{y}$, the Hilbert space is decomposed into a set of superselection sectors labelled by eigenvalues of all $V\left(\xarg\right)$ operators. 
We shall focus here on the sector whose states $\left|\psi\right\rangle$ satisfy
\begin{equation}
    V\left(\xarg\right)\ket{\psi} = \ket{\psi}, \quad \forall \xarg.
    \label{eq:gaugeinv}
\end{equation}

The $N \rightarrow \infty$ limit reproduces compact QED - the Kogut-Susskind Hamiltonian~\cite{kogut_hamiltonian_1975} of a $U(1)$ lattice gauge theory~\cite{wilson_confinement_1974,kogut_introduction_1979}. 
With this analogy in mind, we refer to the first (link) term of~\eqref{eq:hamilton_ZN} as the electric energy and to the second (plaquette) term as the magnetic one.

In this work, we focus on the $N=2$ case - a pure $\mathbb{Z}_2$ lattice gauge theory \cite{wegner_duality_1971,fradkin_order_1978}. There, we can make the choices $Q=Q^{\dagger}=\sigma_x$ and $P=P^{\dagger}=\sigma_z$, and identify $\left\{\left|p\right\rangle\right\}_{p=0}^1$ ($\left\{\left|q\right\rangle\right\}_{q=0}^1$) as the eigenstates of $\sigma_z$ ($\sigma_x$) with eigenvalues $\left(-1\right)^p$ ($\left(-1\right)^q$). The Hamiltonian~\eqref{eq:hamilton_ZN} then simplifies to
\begin{equation} 
    \begin{aligned}
        H &= \lambda\sum_\ell \left[1-P_\ell\right] + \frac{1}{\lambda}\underset{p}{\sum}\left[1-Q_{p_1} Q_{p_2} Q_{p_3}Q_{p_4}\right].
    \end{aligned} 
    \label{eq:ham_gauge}
\end{equation}
(again following the conventions of Fig.~\ref{fig:lattice}).

Traditionally, a link with $\left|p=0\right\rangle$ is considered as one that does not carry any \emph{electric flux} while a link with $\left|p=1\right\rangle$ carries it. $\mathbb{Z}_2$ gauge invariance, in the sector defined by Eq.~\ref{eq:gaugeinv}, implies that we only discuss superpositions of products of $P$ eigenstates, involving only \emph{closed flux loops}.

To gain additional insight into the physics of the model, let us consider the extreme coupling limits. First, the strong coupling limit, where $\lambda \gg 1$, for which it will be convenient to redefine the Hamiltonian as 
$\tilde{H}=\lambda^{-1} H$, and in which the electric term dominates whiile the magnetic term is a small perturbation. For $\lambda \rightarrow \infty$, the ground state takes the form
\begin{equation}
\left|\psi\left(\lambda=\infty\right)\right\rangle = \underset{\ell}{\bigotimes}\left|p=0\right\rangle_{\ell} \equiv \left|\psi_E\right\rangle
\label{eq:psiEdef}
\end{equation}
- a product state of $\left|p=0\right\rangle$ on all the links. If we decrease $\lambda$ a little bit, such that the condition $\lambda \gg 1$ is still satisfied, using perturbation theory one obtains that the ground state becomes 
\begin{equation}
\left|\psi\left(\lambda\gg1\right)\right\rangle = \left(1 + \frac{1}{8\lambda^2}\underset{p}{\sum}Q_{p_1} Q_{p_2} Q_{p_3}Q_{p_4}\right) \left|\psi_E\right\rangle + O\left(\lambda^{-4}\right).
\label{eq:strcoupl}
\end{equation}
The first order correction is a superposition of all the $P$ eigenstates which contain a single plaquette's flux loop; the second order correction will be a superposition of all the possible two-plaquette excitations, and so on. Consider the Wilson loop operator~\cite{wilson_confinement_1974}, which in the $\mathbb{Z}_2$ case takes the form
\begin{equation}
\mathcal{W}\left(\mathcal{C}\right) = \underset{\ell \in \mathcal{C}}{\prod}Q_{\ell}
\label{eq:def_wilson_loop}
\end{equation}
where $\mathcal{C}$ is some close curve on the lattice, usually chosen to be rectangular. 
For a rectangular Wilson loop with dimensions $R_1 \times R_2$, the leading order of the expectation value of the Wilson loop in the strong coupling regime will be given by the $A=R_1 R_2$-th order of the perturbative series of Eq.~\eqref{eq:strcoupl}, and hence the Wilson loop will decay with an \emph{area law},
\begin{equation}
\left\langle \mathcal{W}\left(R_1,R_2,\right)\right\rangle_{\lambda \gg 1} \propto e^{-2\ln \lambda R_1 R_2}
\end{equation}
- a manifestation of a \emph{confining phase}~\cite{wilson_confinement_1974,fradkin_order_1978}.

On the other hand, if we take the other extreme limit of $\lambda \ll 1$, the magnetic energy dominates. 
In the extreme case of $\lambda=0$ the ground state is the magnetic one, $\ket{\psi_B}$, also known as one of the \emph{toric code ground states}~\cite{kitaev_fault-tolerant_2003}. 
It takes the form
\begin{equation}
\left|\psi\left(\lambda=0\right)\right\rangle \propto 
\underset{p}{\prod}\left(1+Q_{p_1} Q_{p_2} Q_{p_3}Q_{p_4}\right) \left|\psi_E\right\rangle \equiv \left|\psi_B\right\rangle
\label{eq:psiB}.
\end{equation}
An equal superposition of all the $P$ states satisfies the gauge invariance condition~\eqref{eq:gaugeinv}. 
Raising $\lambda$ but staying in the $\lambda \ll 1$ regime, we can again use perturbation theory to construct the state, and a straightforward calculation shows that 
\begin{equation}
\left\langle \mathcal{W}\left(R_1,R_2,\right)\right\rangle_{\lambda \ll 1} \propto e^{-2 f\left(\lambda\right) \left(R_1 + R_2\right)}
\end{equation}
- a perimeter law decay, manifesting a \emph{deconfined phase}~\cite{wilson_confinement_1974,fradkin_order_1978}.

The area law and perimeter law extend beyond the perturbative regimes, until one approaches the confinement-deconfinement phase transition~\cite{fradkin_order_1978,elitzur_phase_1979,horn_hamiltonian_1979,peskin_critical_1980,trebst_breakdown_2007,tupitsyn_topological_2010,gregor_diagnosing_2011}.

\section{Construction of an ansatz state\label{sec:construction}}

We would like to build an ansatz state for a variational search of the ground state of the $\mathbb{Z}_2$ Hamiltonian of Eq.~\eqref{eq:ham_gauge}, for different values of $\lambda$. 
What is required from such an ansatz state?

The state must be gauge invariant, and we would like to use a construction which will allow us to couple, in future work, with fermionic matter. 
We therefore use the construction of Gauged Gaussian Fermionic PEPS~\cite{zohar_fermionic_2015,zohar_projected_2016}. 
Furthermore, as was shown in~\cite{zohar_combining_2018}, one can perform all the relevant calculations for such states rather efficiently using Monte-Carlo.

We use this ansatz state, with the $\mathbb{Z}_2$ gauge group, but without physical matter (dynamical fermions). 
The auxiliary degrees of freedom which shall be used for the construction of the state will nevertheless be fermionic, to allow us to couple to fermions later on and enable the efficient Monte-Carlo computation. 
Additionally, we would like our family of ansatz states to contain the extreme $\lambda \gg 1$ and $\lambda \ll 1$ cases discussed above; ideally, the ansatz will be able to interpolate between them and approximate the ground states in the intermediate regime.

We construct our states as follows: for each site $\mathbf{x}$, we introduce $4F$ auxiliary or virtual fermionic modes, associated with the edges of links intersecting through it -- $F$ modes in each direction. 
We denote their creation operators by $\left\{r^{\dagger}_{\alpha}\left(\mathbf{x}\right)\right\}_{\alpha=1}^{F}$, $\left\{u^{\dagger}_{\alpha}\left(\mathbf{x}\right)\right\}_{\alpha=1}^{F}$, $\left\{l^{\dagger}_{\alpha}\left(\mathbf{x}\right)\right\}_{\alpha=1}^{F}$, $\left\{d^{\dagger}_{\alpha}\left(\mathbf{x}\right)\right\}_{\alpha=1}^{F}$, representing the right, up, left and down directions, respectively. 
We will unite the $4F$ virtual modes on a single vertex $\xarg$ under a general notation of the form $\left\{a^{\dagger}_{\alpha}\left(\mathbf{x}\right)\right\}_{\alpha=1}^{4F}$, and define the Gaussian operator
\begin{equation}
A\left(\mathbf{x}\right) = \exp\left(T_{\alpha\beta}a^{\dagger}_{\alpha}\left(\mathbf{x}\right)a^{\dagger}_{\beta}\left(\mathbf{x}\right)\right)
\end{equation}
on each site (in general, $T_{\alpha\beta}$ may be site dependent, but we make it uniform to enforce translation invariance).

We define the virtual symmetry operator $\mathcal{V}\left(\mathbf{x}\right)=\mathcal{V}^{\dagger}\left(\mathbf{x}\right)$ by
\begin{equation}
\mathcal{V}\left(\mathbf{x}\right) a^{\dagger}_{\alpha}\left(\mathbf{x}\right) 
\mathcal{V}^{\dagger}\left(\mathbf{x}\right) = -a^{\dagger}_{\alpha}\left(\mathbf{x}\right)
\end{equation}
and note that
\begin{equation}
    \mathcal{V}\left(\mathbf{x}\right) A\left(\mathbf{x}\right) 
    \mathcal{V}^{\dagger}\left(\mathbf{x}\right) = A\left(\mathbf{x}\right)\, .
    \label{eq:Ainv}
\end{equation}
We can decompose the symmetry operator to
\begin{equation}
    \mathcal{V}\left(\mathbf{x}\right) = \mathcal{V}_r\left(\mathbf{x}\right)\mathcal{V}_u\left(\mathbf{x}\right)\mathcal{V}_l\left(\mathbf{x}\right)\mathcal{V}_d\left(\mathbf{x}\right)
\end{equation}
such that 
\begin{equation}
    \mathcal{V}_r\left(\mathbf{x}\right) r^{\dagger}_{\alpha}\left(\mathbf{x}\right) 
    \mathcal{V}_r^{\dagger}\left(\mathbf{x}\right) = -r^{\dagger}_{\alpha}\left(\mathbf{x}\right)
\end{equation}
and similarly for $u,l,d$.

We further define the gauging operators $\mathcal{U}_G\left(\mathbf{x},i\right)$ (for $i \in \{1,2\}$) on the links, which multiply the fermionic creation (or annihilation) operators of virtual fermions associated with them with the physical gauge field operators $Q$ on the respective links:
\begin{equation}
    \begin{aligned}
        &\mathcal{U}_G\left(\mathbf{x},1\right) r^{\dagger}_{\alpha}\left(\mathbf{x}\right)    
        \mathcal{U}^{\dagger}_G\left(\mathbf{x},1\right) = Q\left(\mathbf{x},1\right)r^{\dagger}_{\alpha}\left(\mathbf{x}\right),    \\
        &\mathcal{U}_G\left(\mathbf{x},2\right) u^{\dagger}_{\alpha}\left(\mathbf{x}\right)    
        \mathcal{U}^{\dagger}_G\left(\mathbf{x},2\right) = Q\left(\mathbf{x},2\right)u^{\dagger}_{\alpha}\left(\mathbf{x}\right).
    \end{aligned}
    \label{eq:UGdef}
\end{equation}
Note that
\begin{equation}
    \begin{aligned}
        P\left(\mathbf{x},1\right) \mathcal{U}_G\left(\mathbf{x},1\right) P^{\dagger}\left(\mathbf{x},1\right)
        &=
        \mathcal{V}_r\left(\mathbf{x}\right) \mathcal{U}_G\left(\mathbf{x},1\right) \mathcal{V}^{\dagger}_r\left(\mathbf{x}\right), \\
        P\left(\mathbf{x},2\right) \mathcal{U}_G\left(\mathbf{x},2\right) P^{\dagger}\left(\mathbf{x},2\right)
        &=
        \mathcal{V}_u\left(\mathbf{x}\right) \mathcal{U}_G\left(\mathbf{x},2\right) \mathcal{V}^{\dagger}_u\left(\mathbf{x}\right)\, .
    \end{aligned}
    \label{eq:gauginginv}
\end{equation}

We also define, on each link, the Gaussian operators
\begin{equation}
    \begin{aligned}
        w\left(\mathbf{x},1\right) &= \exp\left(W^{1}_{\alpha\beta}l^{\dagger}_{\alpha}\left(\mathbf{x}+\hat{\mathbf{e}}_1\right)
        r^{\dagger}_{\beta}\left(\mathbf{x}\right)\right)\\
        w\left(\mathbf{x},2\right) &= \exp\left(W^{2}_{\alpha\beta}d^{\dagger}_{\alpha}\left(\mathbf{x}+\hat{\mathbf{e}}_2\right)
        u^{\dagger}_{\beta}\left(\mathbf{x}\right)\right)
    \end{aligned}
    \label{eq:wdef}
\end{equation}
which couple the virtual fermions associated on both sides of a link and satisfy
\begin{equation}
    \begin{aligned}
        \mathcal{V}_r\left(\mathbf{x}\right) w\left(\mathbf{x},1\right) \mathcal{V}^{\dagger}_r\left(\mathbf{x}\right)&=\mathcal{V}_l\left(\mathbf{x}+\hat{\mathbf{e}}_1\right) w\left(\mathbf{x},1\right) \mathcal{V}^{\dagger}_l\left(\mathbf{x}+\hat{\mathbf{e}}_1\right) \\
        \mathcal{V}_u\left(\mathbf{x}\right) w\left(\mathbf{x},2\right) \mathcal{V}^{\dagger}_u\left(\mathbf{x}\right)&=\mathcal{V}_d\left(\mathbf{x}+\hat{\mathbf{e}}_2\right) w\left(\mathbf{x},2\right) \mathcal{V}^{\dagger}_d\left(\mathbf{x}+\hat{\mathbf{e}}_2\right).
    \end{aligned}
\label{eq:winv}
\end{equation}
The coupling between neighboring sites is necessary because the states would remain a product state otherwise.

Finally, the fermionic Fock vacuum $\left|\Omega\right\rangle$ is invariant under all the virtual  $\mathcal{V}\left(\mathbf{x}\right)$, and the gauge field state $\left|\psi_E\right\rangle$ defined in Eq.~\eqref{eq:psiEdef} is invariant under all the physical gauge transformations $V\left(\mathbf{x}\right)$. 
With all the above definitions, we are ready to define our ansatz state, the fermionic gauged Gaussian PEPS as
\begin{equation}
    \ket{\psi} = \bra{\Omega}
    \underset{\mathbf{x},i}{\prod}w^{\dagger}\left(\mathbf{x},i\right)
    \mathcal{U}_G
    \underset{\mathbf{x}}{\prod}A\left(\mathbf{x}\right)\ket{\Omega}\ket{\psi_E}
    \label{eq:psidef}
\end{equation}
where $\mathcal{U}_G=\underset{\mathbf{x},i}{\prod}\mathcal{U}_G\left(\mathbf{x},i\right)$.
Let us examine this state carefully, to understand what it may describe.

First, on the right, we begin with the physical state $\left|\psi_E\right\rangle$, which is the no-flux, strong coupling limit ground state, as defined in Eq.~\eqref{eq:psiEdef}. 
On top of it, we act with the gauge field operator $\mathcal{O} 
\equiv \bra{\Omega}
    \underset{\mathbf{x},i}{\prod}w^{\dagger}\left(\mathbf{x},i\right)
    \mathcal{U}_G
    \underset{\mathbf{x}}{\prod}A\left(\mathbf{x}\right)\ket{\Omega}$.
Note that this expression contains a part that acts on the gauge-fields that is not traced out.
While all virtual fermions are traced out, the expression still acts as an operator on the gauge fields.
To add physical fermions to the game, one can add their Fock vacuum on the right, and add them to the $A\left(\mathbf{x}\right)$ operator appropriately~\cite{zohar_combining_2018}. 
Here we only focus on the pure gauge case and shall now show how to tailor it to the  requirements mentioned above.
Note that both $
\ket{\psi_E}$ and $\mathcal{O}$ are gauge invariant (this can be shown using the symmetry properties of Eqs. (\ref{eq:Ainv},\ref{eq:gauginginv},\ref{eq:winv})) and therefore our ansatz is gauge invariant, that is, it satisfies 
 Eq.~\eqref{eq:gaugeinv}.

The operator $\underset{\mathbf{x}}{\prod}A\left(\mathbf{x}\right)$ creates a virtual fermionic state when acting on the vaccuum $\ket{\Omega}$. 
It is a product state of the different sites $\mathbf{x}$; on each $\mathbf{x}$, depending on the parameters of $T$, we may excite some, none, or all the virtual modes associated with the four links around it. 
A leg with an odd number of excitations will be referred to as one which carries virtual flux, and one with an even number of excitations will not. 
On top of that, we act with the gauging operator $\mathcal{U}_G$, and then apply the resulting operator to the initial physical state $\ket{\psi_E}$. 
The gauging operator multiplies each creation operator on the outgoing links (right and up) by the physical $Q$ operator on the same link, and thus (since $Q^2=1$), links that carry virtual flux will now also carry a physical one. 
However, the state is still not gauge invariant, since when we take the product of all the sites we might well end up with open flux strings which violate the gauge symmetry; the projection onto $\underset{\mathbf{x},i}{\prod}w^{\dagger}\left(\mathbf{x},i\right)$ prevents that possibility, and ensures that the fluxes are properly connected.

We would like to obtain an ansatz PEPS $\ket{\psi}$ with the minimal $F$ possible, which will include the extreme case ground states, $\ket{\psi_E}$ and $\ket{\psi_B}$, defined in Eq.~\eqref{eq:psiEdef} and~\eqref{eq:psiB} respectively. 
We would also like the state to have rotational invariance, in the lattice sense (invariance under $\pi/2$ rotations). 
In  Appendix~\ref{app:parametrization} we show that this can be obtained for $F=2$ (two virtual fermions on each link) with 
\begin{align}
  T=\left(
  \begin{array}{ccccccccc}
    & 0     & -z_1  & -i y_1  & -i z_1 & i a   & i b  & i c    & i d   \\
    & z_1    & 0    & -i z_1  & y_1    & -d    & -a   & -b     & -c    \\
    & i y_1  & i z_1 & 0      & z_1    & -i c  & -i d & -i a   & -i b  \\
    & i z_1  & -y_1  & -z_1    & 0     & b     & c    & d      & a     \\
    & -i a  & d    & i c    & -b    & 0     & -z_2  & -i y_2  & -i z_2 \\
    & -i b  & a    & i d    & -c    & z_2    & 0    & -i z_2  & y_2    \\
    & -i c  & b    & i a    & -d    & i y_2  & i z_2 & 0      & z_2    \\
    & -i d  & c    & i b    & -a    & i z_2  & -y_2  & -z_2    & 0     \\
  \end{array}
  \right) \, .
  \label{eq:tmat_rotation_approach_ordered}
\end{align}
(where the ordering is $r_1,u_1,l_1,d_1,r_2,u_2,l_2,d_2$ and all the parameters are, in general, complex)
and
\begin{equation}
W^1 = \sigma_x,\quad W^2 = \eta^2 \sigma_x
\label{eq:Wmat}
\end{equation}
(where $\eta^4=-1$, and this parameterization was derived for the choice $\eta = e^{i\pi/4}$ - equivalent parameterizations for other possible values exist too).

As we show in the appendix, setting all the parameters to zero gives rise to $\ket{\psi_E}$; setting them all to zero besides $b^4=-\frac{1}{16}$ produces $\ket{\psi_B}$. 

A gauge invariant construction can be obtained as well with one virtual fermion per link, $F=1$, and (as can be seen in the appendix) it captures the physics for large and small values of the coupling constant $\lambda$ with good precision, though not exactly. We refer to this construction as the \enquote{minimal} ansatz. However, as can be seen numerically, it fails in the intermediate coupling regime. Therefore, in the main text
we focus on the $F=2$ (cf. Eq.~\eqref{eq:tmat_rotation_approach_ordered}) approach, which we denote the \enquote{optimized} ansatz.
For further details and a direct comparison of the two approaches, we refer to Appendix~\ref{app:parametrization}.

\section{The Algorithm\label{sec:algorithm}}
Having introduced and justified the ansatz physically, we would now like to recall why it is useful numerically and discuss the computation algorithm. 
We tackle this problem in two steps: first we show how to compute expectation values for a given set of parameters by combining GGFPEPS with Monte Carlo~\cite{zohar_combining_2018}.
We then demonstrate how to adapt the parameters using variational Monte Carlo (VMC) methods.

\subsection{$Q$ eigenbasis formulation}

Following~\cite{zohar_combining_2018}, we begin by expressing our ansatz PEPS $\ket{\psi}$ in the $Q$ eigenbasis. 
We introduce the gauge field configuration states, which are simply product states of $Q$ eigenstates on all the lattice's links:
\begin{equation}
    \ket{\mathcal{Q}}=\underset{\mathbf{x},i}{\bigotimes}\ket{q\left(\mathbf{x},i\right)}.
\end{equation}
As eigenstates of all the $Q$ operators, they satisfy
\begin{equation}
    Q\left(\mathbf{x},i\right)\ket{\mathcal{Q}}=e^{i\pi q\left(\mathbf{x},i\right)}\ket{\mathcal{Q}}.
\end{equation}
Due to the orthonormality of the local $\ket{q}$ state on each link, these states are also orthonormal:
\begin{equation}
    \braket{\mathcal{Q}}{\mathcal{Q}'}=\delta_{\mathcal{Q},\mathcal{Q}'}=
    \underset{\mathbf{x},i}{\prod}\delta_{q\left(\mathbf{x},i\right),q'\left(\mathbf{x},i\right)}.
\end{equation}

In this basis, the gauging operators $\mathcal{U}_G\left(\mathbf{x},i\right)$ from Eq.~\eqref{eq:UGdef} may be seen as controlled operations, transforming the fermionic operators based on the gauge field's $Q$ eigenvalue,
\begin{equation}
    \mathcal{U}_G\left(\mathbf{x},i\right) =  \underset{q}{\sum}\ket{q}_{\mathbf{x},i}\bra{q}_{\mathbf{x},i} \otimes \mathcal{U}_q\left(\mathbf{x},i\right)
\end{equation}
where
\begin{equation}
    \begin{aligned}
        \mathcal{U}_q\left(\mathbf{x},1\right) &= \exp\left(i\pi q \overset{F}{\underset{\alpha=1}{\sum}}r^{\dagger}_{\alpha}\left(\mathbf{x}\right)r_{\alpha}\left(\mathbf{x}\right)\right)\\
        \mathcal{U}_q\left(\mathbf{x},2\right) &= \exp\left(i\pi q \overset{F}{\underset{\alpha=1}{\sum}}u^{\dagger}_{\alpha}\left(\mathbf{x}\right)u_{\alpha}\left(\mathbf{x}\right)\right).
    \end{aligned}
\end{equation}
As a result (and up to an irrelevant normalization factor - the PEPS is not normalized in any case), we can rewrite the ansatz state as
\begin{equation}
\ket{\psi}=\underset{\mathcal{Q}}{\sum}\psi\left(\mathcal{Q}\right)\left|\mathcal{Q}\right\rangle.
\label{eq:psiQbasis}
\end{equation}
The wavefunction is given by
\begin{equation}
    \psi\left(\mathcal{Q}\right) = 
    \bra{\Omega}\underset{\mathbf{x},i}{\prod}w^{\dagger}\left(\mathbf{x},i\right)
    \mathcal{U}_{\mathcal{Q}}
    \underset{\mathbf{x}}{\prod}A\left(\mathbf{x}\right)\ket{\Omega}
    \label{eq:wavefunctiondef}
\end{equation}
where
\begin{equation}
\mathcal{U}_{\mathcal{Q}} = \underset{\mathbf{x},i}{\prod} \mathcal{U}_{q}\left(\mathbf{x},i\right).
\end{equation}

The wavefunction $\psi\left(\mathcal{Q}\right)$ is nothing but an overlap of two fermionic Gaussian states, 
\begin{equation}
    \begin{aligned}
        &\ket{\psi_R} = \underset{\mathbf{x}}{\prod}A\left(\mathbf{x}\right)\ket{\Omega}\\
        &\ket{\psi_L\left(\mathcal{Q}\right)} = \mathcal{U}_{\mathcal{Q}} 
        \underset{\mathbf{x},i}{\prod}w\left(\mathbf{x},i\right)\ket{\Omega}
    \end{aligned}
\end{equation}
(we have used the fact that $\mathcal{U}_{\mathcal{Q}} =\mathcal{U}_{\mathcal{Q}}^{\dagger}$). 

Fermionic Gaussian states are fully classified by the elements of their covariance matrices~\cite{bravyi_lagrangian_2005}; the covariance matrix consists of correlators of products of two fermionic operators (fermionic two-point, or Green's, functions). 
They are central in our algorithm, based on conventional Gaussian fermionic PEPS techniques~\cite{kraus_fermionic_2010}. 

So far, we have expressed the fermionic modes using creation and annihilation (Dirac) operators. 
Numerically and analytically, however, following the Gaussian formalism discussed in~\cite{bravyi_lagrangian_2005} and in the context of PEPS in~\cite{kraus_fermionic_2010}, it is more convenient to work with Majorana modes since they yield real-valued covariance matrices.
As usual, for a given Dirac mode annihilated by $c$ and created by $c\dgr$, we define two Majorana modes
\begin{align}
    \begin{aligned}
        \maj{\gamma}{1}&=c+c\dgr \\
        \maj{\gamma}{2}&=i(c-c\dgr)
    \end{aligned}
    \label{eq:app_def_majorana_modes}
\end{align}
- that is, a system with $p$ Dirac modes has $2p$ Majorana modes, labeled from $1$ to $2p$.
The Majorana modes anti-commutation relations are given by the Clifford algebra 
\begin{equation}
\acomm{\gamma_i}{\gamma_j}=2\delta_{ij}.   
\end{equation}

The covariance matrix of a state $\ket{\Phi}$ is given by
\begin{align}
\label{eq:def_covariance_maj}
  \Gamma_{a,b}&=\frac{i}{2}\expval{\left[\gamma_a,\gamma_b\right]}\\
  &=\frac{i}{2}\frac{\expval{\left[ \gamma_a,\gamma_b \right]}{\Phi}}{\braket{\Phi}}.
\end{align}
In Appendix~\ref{app:covariance_matrices} we show how to compute the covariance matrix of a state expressed in terms of exponentials of creation operator bilinears acting on the Fock vacuum, as the states $\ket{\psi_R}$ and $\ket{\psi_L\left(\mathcal{Q}\right)}$ that we work with.

We denote the covariance matrix of $\ket{\psi_R}$ by $D$, and that of $\ket{\psi_L\left(\mathcal{Q}\right)}$ by $\Gamma_{\text{in}}\left(\mathcal{Q}\right)$. 
Since both $\ket{\psi_R}$ and $\ket{\psi_L\left(\mathcal{Q}\right)}$ are product states, it is easy to construct their covariance matrices out of local ingredients: both covariance matrices will be block diagonal. $D$ will be a direct sum of identical blocks, each being the covariance matrix of the state created by $A\left(\mathbf{x}\right)$ on a single site; $\Gamma_{\text{in}}\left(\mathcal{Q}\right)$ is a direct sum of covariance matrices of pairs of virtual fermions on the links, which will not be identical due to the gauging.

\subsection{The norm of the state}
All our computations will be based on computing expectation values of operators with respect to the ansatz state $\ket{\psi}$, in the form of Eq.~\eqref{eq:psiQbasis}. 
However, as this state is not normalized, we would like to show how the norm is computed.
Note that
\begin{equation}
    \braket{\psi}{\psi} = \underset{\mathcal{Q}}{\sum}\left|\psi\left(\mathcal{Q}\right)\right|^2
    =\Tr\left[\ket{\psi_R}\bra{\psi_R}\ket{\psi_L\left(\mathcal{Q}\right)}\bra{\psi_L\left(\mathcal{Q}\right)}\right]
\end{equation}
- the squared norm of $\ket{\psi}$ is nothing but a sum over overlaps between two Gaussian fermionic density matrices. 
This can easily computed in terms of their covariance matrices~\cite{mazza_quantum_2012} giving rise to:
\begin{align}
  \left|\psi\left(\mathcal{Q}\right)\right|^2=\sqrt{\det(\frac{1-\gammain\left(\mathcal{Q}\right) D}{2})}.
  \label{eq:overlap_default}
\end{align}

\subsection{Computation of the Wilson Loop}
The first expectation value we are interested in computing is that of a Wilson loop $\mathcal{W}\left(\mathcal{C}\right)$ as defined in 
Eq.~\eqref{eq:def_wilson_loop}.

The key property here \cite{zohar_combining_2018} is that the configuration states $\ket{\mathcal{Q}}$ are eigenstates of the $Q$ operators and thus also of the Wilson loop operator:
\begin{equation}
    \mathcal{W}\left(\mathcal{C}\right)\ket{\mathcal{Q}}=\underset{\ell\in\mathcal{C}}{\prod}\left(-1\right)^{q\left(\ell\right)}\ket{\mathcal{Q}}
\end{equation}
where the $q\left(\ell\right)$ values are dictated by the eigenvalues of $Q_{\ell}$ operators with respect to the configuration of $\ket{\mathcal{Q}}$. Therefore,
\begin{align}
  \expval{\mathcal{W}(\mathcal{C})}=\sum_\gauge F_{W(\mathcal{C})}(\gauge)p(\gauge),
  \label{eq:exp_obs_wilson}
\end{align}
where $F_{W(C)}=\underset{\ell\in\mathcal{C}}{\prod}\left(-1\right)^{q\left(\ell\right)}$ and we define the function
\begin{align}
  p(\gauge)=\frac{\abs{\Psi(\gauge)}^2}{\sum_{\gauge'}\abs{\Psi(\gauge')}^2}.
  \label{eq:def_prob}
\end{align}
Note that for any $\mathcal{Q}$, $0 \leq p\left(\mathcal{Q}\right)\leq 1$, and that $\underset{\mathcal{Q}}{\sum}p\left(\mathcal{Q}\right)=1$ and thus it is a probability density function over the gauge field configuration space.

Using Metropolis sampling~\cite{metropolis_equation_1953}, we can compute the expectation value using Markov chain Monte Carlo sampling (MCMC) \cite{zohar_combining_2018}.
Instead of the full probability $p(\gauge)$, we need only the transition probability between two gauge field configurations,
\begin{align}
    p(\gauge \rightarrow \gauge') = \frac{\abs{\Psi(\gauge')}^2}{\abs{\Psi(\gauge)}^2}\, .
\end{align}
The denominator of Equation~\eqref{eq:def_prob}, which is hard to compute, is avoided.
In the Monte Carlo procedure, we use a single-site update, i.e. we randomly select a single site and propose a new gauge field for it.
Since the changes in the covariance matrices are only local, we can use the matrix-determinant lemma and the Woodbury identity to update inverses and determinants locally after each step.

In general, the exact contraction of a general PEPS is exponentially hard~\cite{schuch_computational_2007}.
Here, since we picked the subclass of  gauged Gaussian fermionic PEPS, we can perform the contraction required for Wilson loop computation efficiently, using covariance matrices and Eq.~\eqref{eq:overlap_default}.

\subsection{Computing $P$ expectation values\label{sec:p_exp_value}}
The next operator whose expectation value we would like to compute is $P$ on a given link. 
It does not act diagonally on the configuration states $\ket{Q}$, and thus this has be done with caution~\cite{zohar_combining_2018}.

First, note that for a given link $\ell$,
\begin{equation}
    \bra{\psi}P_{\ell}\ket{\psi}=
    \underset{\mathcal{Q},\mathcal{Q}'}{\sum}
    \overline{\psi\left(\mathcal{Q}'\right)}\psi\left(\mathcal{Q}\right)\bra{\mathcal{Q}'}P_{\ell}\ket{\mathcal{Q}}.
\end{equation}
$P_{\ell}$ changes $q$ on the link $\ell$ and does not affect any other links; thus, the configuration $\mathcal{Q}'$ is identical to $\mathcal{Q}$ everywhere but on $\ell$, where we have the opposite $q$ eigenvalue. 
If we denote $\hat{Q}$ as the configuration of gauge fields on all the links but $\ell$, such that $\mathcal{Q}=(\hat{Q},q)$ and $\mathcal{Q}'=(\hat{Q},q-1)$ (the subtraction operation is obviously modulo $2$), we have
\begin{equation}
    \begin{aligned}
        \left\langle P_{\ell} \right\rangle =\frac{\bra{\psi}P_{\ell}\ket{\psi}}{\braket{\psi}{\psi}}&=
        \frac{\underset{\mathcal{Q}}{\sum}
        \overline{\psi\left(\hat{Q},q-1\right)}
        \psi\left(\hat{Q},q\right)}
        {\underset{\mathcal{Q}'}{\sum}\left|\psi\left(\mathcal{Q}'\right)\right|^2}\\
        &=\underset{\mathcal{Q}}{\sum}
        \frac{\overline{\psi\left(\hat{Q},q-1\right)}
        \psi\left(\hat{Q},q\right)}{\left|\psi\left(\mathcal{Q}\right)\right|^2}
        p\left(\mathcal{Q}\right)\\
        &\equiv\underset{\mathcal{Q}}{\sum}F_P\left(\mathcal{Q}\right)p\left(\mathcal{Q}\right).
    \end{aligned}
    \label{eq:Pexp}
\end{equation}
Therefore, if we have an efficient way to compute $F_P\left(\mathcal{Q}\right)$, we can use Monte-Carlo techniques to evaluate the expectation value of $P$ as well. 
In Appendix~\ref{app:electric_energy} we show how this can be done.

In a previous work~\cite{emonts_variational_2020}, the electric energy was calculated by explicitly transforming the expression to Grassmann variables.
The resulting equation contains a Pfaffian that depends on the system-size.
In contrast to the determinants and inverses that are used in the algorithm, the value of the Pfaffian cannot be tracked across Monte Carlo updates.
Thus, the computation of a system-sized Pfaffian is necessary with each measurement.

In this paper, we introduce a new way to compute the electric energy that only depends on Pfaffians of constant size if $F>1$.
For $F=1$, the computation does not depend on Pfaffians at all (cf. Appendix~\ref{app:electric_energy}).
Here, we use the properties of the Gaussian mapping for covariance matrices to obtain the numerical value of the electric energy.
In the case of the optimized ansatz ($F=2$), Pfaffians of constant size enter the computation, but since they do not scale with system size, they do not hamper the computation.

\subsection{Looking for the ground state}
We wish to find the ground state by minimizing the expectation value of the Hamiltonian given in Eq.~\eqref{eq:ham_gauge}. 
Thanks to the translational and rotational invariance of the Hamiltonian and our ansatz, we can express the energy to be minimized as
\begin{align}
  E=n_l \lambda \left(1-\expval{P}\right)+n_p \left( 1- \expval{Q_{1}Q_{2}Q_{3}Q_{4}}\right)
  \label{eq:hamilton_trans_inv}
\end{align}
where $n_l$ is the number of links in the system and $n_p$ is the number of plaquettes, and $\expval{P}$, $\expval{Q_{1}Q_{2}Q_{3} Q_{4}}$ refer to the expectation value on one particular link and plaquette which we can choose arbitrarily.

The Monte Carlo procedure described above enables us to compute these expectation values for a given set of parameters $\alpha$.
In addition to the evaluation, we need a minimization step that drives the parameters towards the groundstate.

By computing the gradient of the energy with respect to the parameters $\dv{E}{\alpha}$, we can use the Broyden–Fletcher–Goldfarb–Shanno (BFGS) algorithm ~\cite{press_numerical_2007} to minimize the parameter values.

\section{Results\label{sec:results}}
In section~\ref{sec:construction}, we introduced a gauge invariant ansatz state depending on complex parameters.
Its expressive power depends on the number of virtual fermions $F$ on the links.
In the following section, we will numerically benchmark the state and explicilty compare the minimal construction ($F=1$) with the optimized construction ($F=2$).
Following analytic arguments (cf. Appendix~\ref{app:parametrization}), we expect the optimized ansatz to match better with exact results.

The key part of the numerical computation of the energy is the evaluation of the sum in Eq.~\eqref{eq:psiQbasis}.
In general, the number of terms in the sum scales exponentially with the lattice size.
Thus, for large systems, we cannot expect to evaluate the sum exactly and we resort to Monte Carlo (MC) computations.
For small systems, however, an exact contraction (EC) of the states is feasible.
The exact evaluation on small systems decouples the error that we introduce by sampling with Monte Carlo from problems with the ansatz itself: even if we evaluate a bad ansatz state perfectly with MC, it stays a bad ansatz state.

\begin{figure}
    \centering
    \includegraphics[width=\columnwidth]{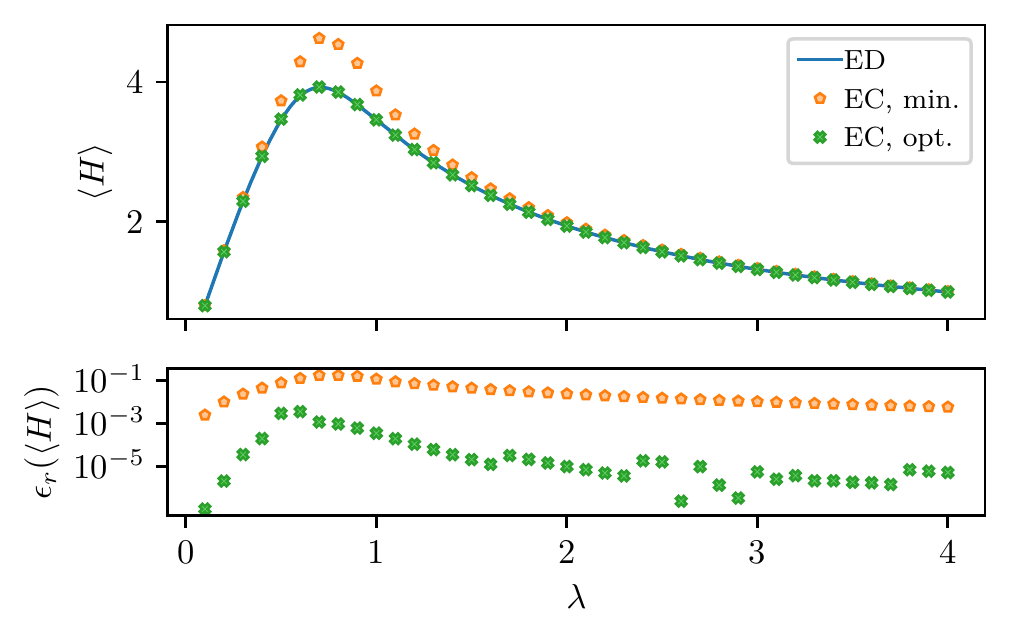}
    \caption{Comparison of exact diagonalization (ED) data for a $2\times 2$ system with exact contraction (EC) data obtained by explicitly contracting the GGFPEPS.
    The upper panel shows a comparison of the minimal and the optimized approach with ED data.
    The lower panel displays the relative error between each of the two approaches with the ED data.
    }
    \label{fig:ec_vs_ed}
\end{figure}

As a first step, we compare the result of an exact contraction GGFPEPS with ED data on a $2\times2$ system (cf. Fig.~\ref{fig:ec_vs_ed}).
The notation $2\times 2$ corresponds to a single plaquette that is closed with periodic boundary conditions, leading to $4$ plaquettes in total.
Here, exact diagonalization refers to solving the time-independent Schrödinger equation explicitly by diagonalizing the Hamiltonian.
This is possible for a $\Z{2}$ gauge theory since the link Hilbert spaces have finite dimension.

For high couplings, where the electric term dominates, the variationally minimized data agrees well with the ED data.
This behavior is expected since the electric ground state (no magnetic term) can be exactly represented with the minimal approach.
We expect that the minimal approach has larger problems in the low-coupling region (dominated by the magnetic energy).
In contrast to earlier works~\cite{zohar_combining_2018,emonts_variational_2020}, however, we see a good agreement at lower couplings as well.
The main difference stems from allowing the parameters in $T$ to be complex.
For a more detailed analysis, we refer to Appendix~\ref{app:parametrization}.

The lower panel of Figure~\ref{fig:ec_vs_ed} shows the relative error $\epsilon_r(\expval{H})=\left(\expval{H}_\text{EC}-\expval{H}_\text{ED}\right)/\expval{H}_\text{ED}$, where the subscripts of the expectation values indicate the method of computation.
The plot illustrates that the convergence of the optimized ansatz is not only better in the transition region, but across the entire coupling region.

\begin{figure}
    \centering
    \includegraphics[width=\columnwidth]{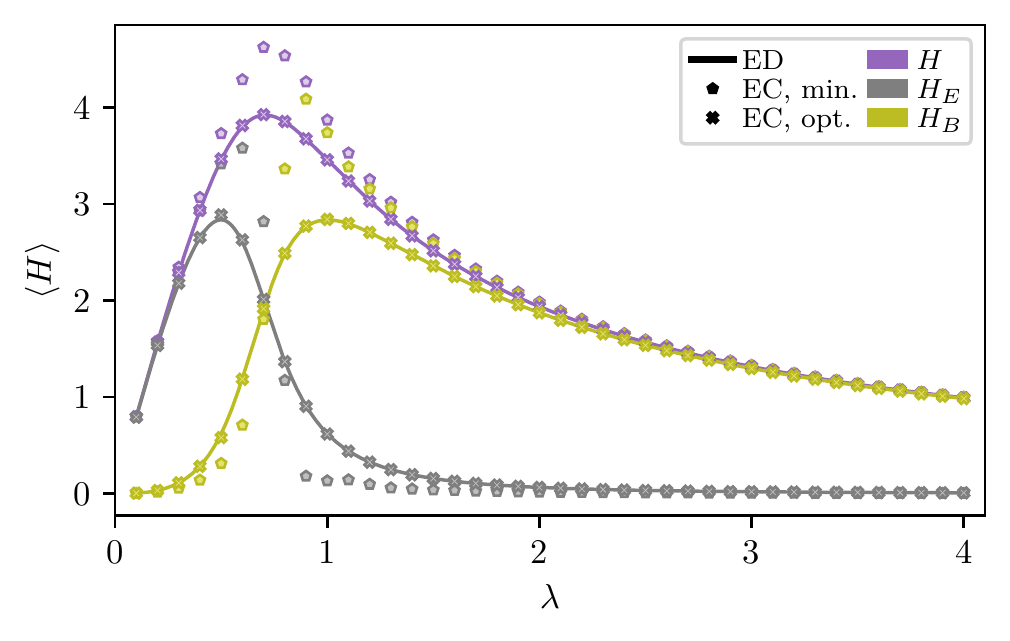}
    \caption{Comparison of different energy components between exact contraction (EC) and exact diagonalization (ED) on a $2\times 2$ lattice.
    The color of the lines and markers encodes the part of the energy: total energy (purple), electric energy (gray), and magnetic energy (olive).
    The shape of the markers indicates the computational method: ED (full line), GGFPEPS with the minimal approach (pentagons), and GGFPEPS with the optimized approach (crosses).}
    \label{fig:partial_energy_comparison}
\end{figure}

Since we know from the analytic considerations above that the groundstate of the magnetic term cannot be exactly represented by the minimal approach, we take a more detailed look at the different terms of the Hamiltonian.
The minimization of the total energy of a system is easier than obtaining the correct results for other observables.
Figure~\ref{fig:partial_energy_comparison} shows exact contraction data of the two approaches for a system of size $2\times 2$.
To further study the problem of the minimal approach, we visualize the total energy $H$, the electric $H_E$ and the magnetic $H_B$ with different colors.
The curves of the total energy are the same as in the upper panel of Fig.~\ref{fig:ec_vs_ed}.
The minimal approach matches decently for the low coupling region and well for the high coupling region. 

In the transition region around $g=0.8$, however, the total energy is far from optimal and the decomposition into electric energy and magnetic energy does not follow the actual groundstate (given by ED) at all.
Note that the gray points (electric energy) and the olive points (magnetic energy) are allowed to lie under the exact solution (solid line).
The variational principle holds only for the total energy and not for individual parts of the total energy.

The data for the optimized approach shows two distinct advantages over the minimal approach.
Firstly, it fits much closer to the exact solution, especially in the transition regions.
Secondly, the variational results follow the magnetic energy and the electric energy much better.
Thus, the optimized approach is a more faithful description of the actual ground state.

The exponential scaling of the number of configurations renders the exact contraction scheme for larger systems extremely computationally expensive.
For these systems, we use Monte Carlo sampling.

All variational Monte Carlo data shown in the plots is obtained with $10^5$ warm-up steps and $10^5$ measurement steps.
Our ansatz is translationally invariant and the number of parameters scales only with $F$ and not with the system size $L^2$ where $L$ is the linear extent of the lattice.
Thus, we can use the results of EC computations as starting points for the Monte Carlo computations of larger system sizes.
The error bars on the variational Monte Carlo data are computed with a re-binning analysis to take into account the finite auto-correlation from the single-site update.

\begin{figure}
    \centering
    \includegraphics[width=\columnwidth]{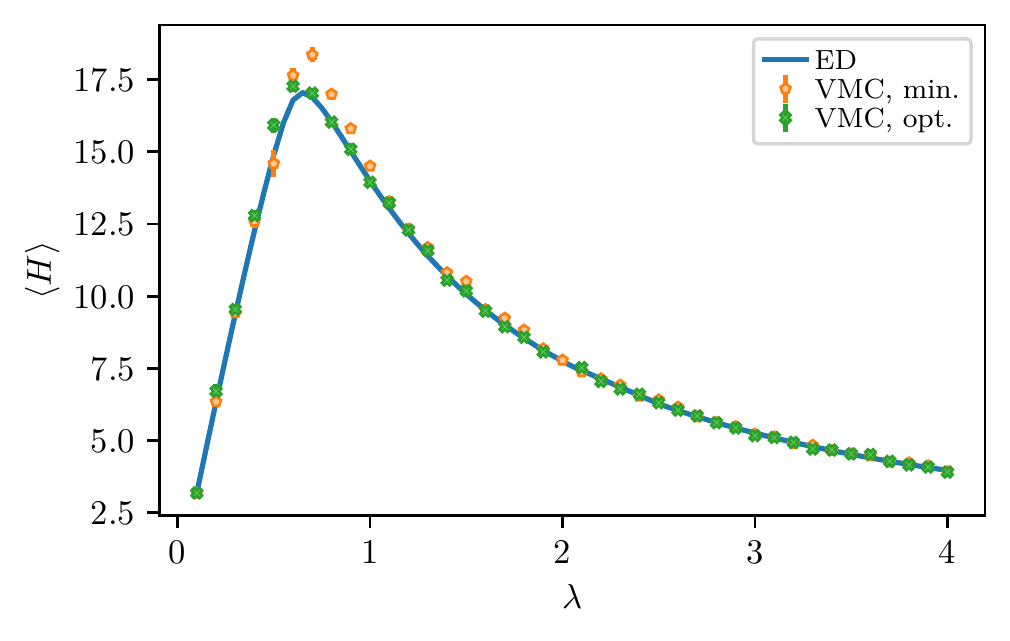}
    \caption{Comparison of exact diagonalization data for a $4\times 4$ system with variational Monte Carlo data for the same systemsize.
    }
    \label{fig:mc_vs_ed}
\end{figure}

In the previous paragraphs, we established that the ansatz is well-suited to describe the physics of $\Z{2}$ lattice gauge theories.
In Fig.~\ref{fig:mc_vs_ed}, we check the variational Monte Carlo sampling procedure.
We compare data for a $4\times 4$ system computed with Monte Carlo sampling for the GGFPEPS and with an exact diagonalization code exploiting the symmetries of the system~\cite{borla_quantum_2022},  which relies on the ED library QuSpin~\cite{weinberg_quspin_2017,weinberg_quspin_2019}.
The data shows good agreement of the Monte Carlo sampling procedure with ED data.

\begin{figure}
    \centering
    \includegraphics[width=\columnwidth]{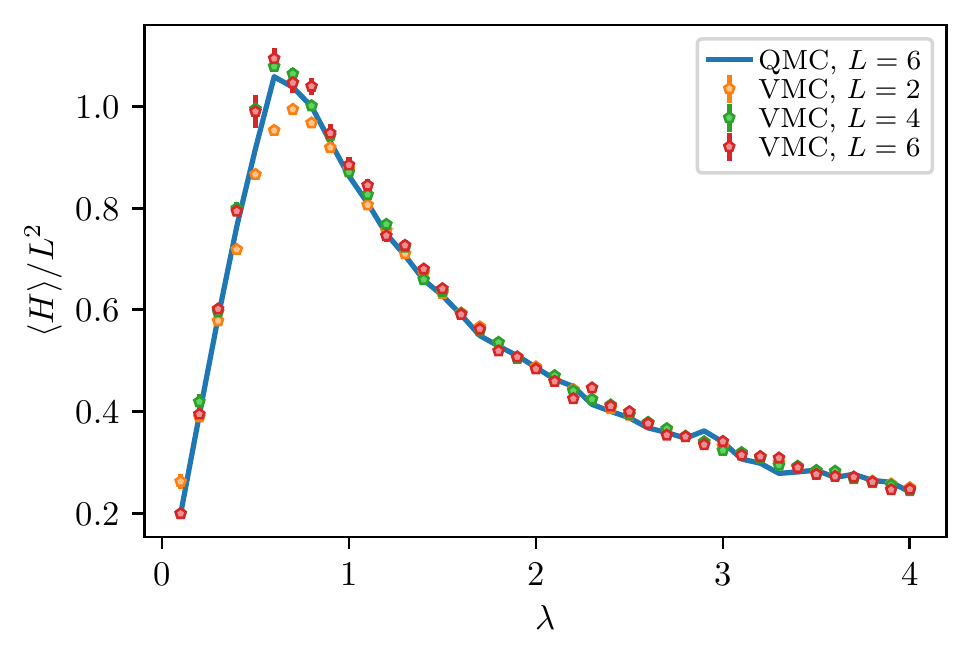}
    \caption{Comparison of VMC data of the optimized ansatz with data obtained via QMC. Data of different system sizes is scaled with the number of vertices $L^2$ where $L$ is the linear extent of the lattice.}
    \label{fig:size_scaling}
\end{figure}

Finally, we can start extending the system to sizes that are not accessible with exact contractions or exact diagonalization.
In Fig.~\ref{fig:size_scaling}, we show the energy of the optimized ansatz ($F=2$) for different sizes.
For better comparison, the energy is scaled to the number of vertices $L^2$ on the lattice.
Additionally, we show data of standard quantum Monte Carlo (QMC) on a $6\times 6$ as a comparison~\cite{gazit_emergent_2017}.
The good agreement between the $L=4$ and $L=6$ data can indicate that finite size effects do not have a large effect on the energy of the system.
The situation might be different for other observables.

\section{Summary and Conclusions\label{sec:conclusions}}
In this paper, we constructed an efficient variational ansatz for the pure \Z{2} lattice gauge Hamiltonian.
The ansatz in form of a GGFPEPS explicitly fulfills the gauge invariance of the system.
The expressibility of the ansatz can be controlled by the number of virtual fermions on the links.
We show analytically that a single virtual fermion on the links ($F=1$) is not sufficient to represent the weak-coupling groundstate of the theory.

Numerically, we check the analytical result by exactly contracting the state and showing that the optimized approach represents the energy more faithfully.
Due to the Gaussian character of the state, the contraction is efficient and larger systems can be handled via Monte Carlo sampling.

Additionally, we demonstrate a new method to compute the electric energy for GGFPEPS.
In previous work~\cite{emonts_variational_2020}, the computation demanded the computation of a system-sized Pfaffian in every measurement.
This computation is substituted by a matrix multiplication and an inversion which can be tracked through the local sampling procedure.
This can enable the exploration of larger systems and removes one of the big runtime penalties of the algorithm.

One of the immediate next steps to add physical fermions to the system.
The formulation of the state is written in terms of fermions to allow a seamless integration of physical fermions in the ansatz.

Furthermore, the more efficient formulation algorithm introduce here, may enable simulations in three space dimensions, as well as of other gauge groups, including compact ones such as $U(1)$, $SU(2)$ or $SU(3)$, and quite possibly serve as a way to study non-perturbative gauge theories, in particular $3+1d$ QCD with a finite chemical potential, which suffers from the sign problem.

\begin{acknowledgements}
E.Z. and P.E. thank Ignacio Cirac for fruitful discussions.
P.E. acknowledges support from the International Max-Planck Research School for Quantum Science and Technology (IMPRS-QST). 
A.K. and E.Z. acknowledge  the support of the Israel Science Foundation
(grant No. 523/20). S.G. acknowledges support from the Israel Science Foundation (grant no. 1686/18). S.M. is supported by
Vetenskapsr\aa det (grant number 2021-03685).
\end{acknowledgements}

\bibliography{references.bib}

\appendix
\section{Parametrization\label{app:parametrization}}

In the main text, the parametrization of the ansatz state is presented in a rather brief manner. Here, we provide more details and complete the required proofs.

\subsection{Rotational Invariance}
First, we would like to ensure that the state is rotationally invariant, in the lattice sense, that is, invariant under $\pi/2$ rotations. We will do so by briefly reviewing the procedures of refs \cite{zohar_fermionic_2015,zohar_projected_2016}. Let us first define how rotations are carried out. Given a lattice site, $\mathbf{x}=\left(x_1,x_2\right)$, we define its rotation by 
\begin{equation}
\Lambda\mathbf{x}=\left(-x_2,x_1\right).
\end{equation}
We can then define the rotation of physical operators simply as
\begin{equation}
\begin{aligned}
Q\left(\mathbf{x},1\right) \rightarrow 
\mathcal{R}^p Q\left(\mathbf{x},1\right) \mathcal{R}^{p\dagger} &= Q\left(\Lambda\mathbf{x},2\right)\\
Q\left(\mathbf{x},2\right) \rightarrow 
\mathcal{R}^p Q\left(\mathbf{x},2\right) \mathcal{R}^{p\dagger} &= Q\left(\Lambda\mathbf{x}-\hat{\mathbf{e}}_1,1\right)
\end{aligned}
\end{equation}
where $\mathcal{R}^p$ is the unitary operator implementing the $\pi/2$ rotation of physical operators and states, and similar relations hold for the $P$ operators. Clearly, the strong coupling vacuum is rotation invariant,
\begin{equation}
\mathcal{R}^p \ket{\psi_E} = \ket{\psi_E},
\end{equation}
and therefore it is easy to see that the weak coupling vacuum $\ket{\psi_B}$ is invariant too, following its definition in Eq.~\eqref{eq:psiB}.

We also define the rotation of the virtual degrees of freedom, implemented by the unitary $\mathcal{R}^v$, by
\begin{equation}
    \mathcal{R}^v a^{\dagger}_{\alpha} \left(\mathbf{x}
    \right) \mathcal{R}^{v\dagger} = R_{\alpha \beta} a^{\dagger}_{\beta} \left(\Lambda\mathbf{x}\right)
\end{equation}
where $R_{\alpha \beta}$ is a matrix which relates to the rotation of the legs: taking $r$ to $u$, $u$ to $l$ etc. Clearly, it has to be unitary. Thus, it has to be a permutation matrix (up to phases). If we choose the ordering
$a^{\dagger}_{\alpha}=
\left(r^{\dagger}_1,u^{\dagger}_1,l^{\dagger}_1,d^{\dagger}_1,...,r^{\dagger}_F,u^{\dagger}_F,l^{\dagger}_F,d^{\dagger}_F\right)^T$ and assume that the rotation does not create any mode mixing, we can simply write $R$ as the direct sum
\begin{equation}
    R= \overset{F}{\underset{m=1}{\bigoplus}}R_0.
\end{equation}
where 
\begin{align}
  R_0&=\eta\mqty(0&1&0&0\\0&0&1&0\\0&0&0&1\\1&0&0&0)
  \label{eq:app_rot_minimal}
\end{align} 
where $\left|\eta\right|=1$. If we wish to couple to physical fermions, the virtual fermions should have similar transformation rules \cite{zohar_fermionic_2015}; since a complete rotation puts a minus sign on a fermion, this phase must satisfy
\begin{equation}
\eta^4=-1.
\label{eq:etadef}
\end{equation}

Note that the virtual vacuum is invariant under rotations,
\begin{equation}
\mathcal{R}^v \ket{\Omega} = \ket{\Omega},
\end{equation}
and thus we can guarantee, following  Refs. \cite{zohar_fermionic_2015,zohar_projected_2016}, that our PEPS $\ket{\psi}$, as defined in Eq.~\eqref{eq:psidef} will be rotationally invariant if
\begin{equation}
    \mathcal{R}^p \mathcal{R}^v A\left(\mathbf{x}\right) \mathcal{R}^{v\dagger} \mathcal{R}^{p\dagger} =A\left(\Lambda\mathbf{x}\right)
    \label{eq:Arot}
\end{equation}
as well as
\begin{equation}
    \begin{aligned}
        \mathcal{R}^v w\left(\mathbf{x},1\right) \mathcal{R}^{v\dagger}  &=w\left(\Lambda\mathbf{x},2\right),\quad\text{and}\\
        \mathcal{R}^v w\left(\mathbf{x},2\right) \mathcal{R}^{v\dagger}  &=w\left(\Lambda\mathbf{x}-\hat{\mathbf{e}}_1,1\right).
    \end{aligned}
    \label{eq:wrot}
\end{equation}

The rotation property of $A\left(\mathbf{x}\right)$ from Eq.~\eqref{eq:Arot} is obtained if and only if the matrix $T$ satisfies the equation
\begin{equation}
    R^T T R = T
    \label{eq:Tpar}
\end{equation}
(just by acting with the rotation operators on the exponential of $A\left(\mathbf{x}\right)$ explicitly). This further constrains the parameterization.

The rotation rules of $w\left(\mathbf{x},i\right)$ may be demanded in a similar fashion. Using the definitions from Eq.~\ref{eq:wdef}, we get that the proper transformation rules~\eqref{eq:wrot} are obtained if and only if
\begin{equation}
    \begin{aligned}
        \eta^2 W^1 &= W^2 \\
        \eta^2 W^2 &= -W^{1T}
        \label{eq:wpar}
    \end{aligned}
\end{equation}
giving rise  the consistency equation
\begin{equation}
    \eta^4 W^1 = -W^{1T}.
\end{equation}
Since we do not wish $\eta$ to depend on $F$, let us consider what happens for $F=1$ where $W^{1,2}$ are simply numbers and their transposition is irrelevant; this forces us to satisfy Eq. (\ref{eq:etadef}) even in the absence of physical fermions.

We choose not to include any free parameters in $W^{1,2}$; a common trick in PEPS theory \cite{cirac_matrix_2021} allows one to absorb all the free parameters into the on-site tensors ($T$ in our case). 

With all that at hand, we can now proceed to construct the most suitable ansatz state $\ket{\psi}$ while meeting the required contstraints.

\subsection{Minimal ansatz}

The most minimal construction we may try is $F=1$: a single virtual fermion per leg.
Then, $T_{\alpha \beta}$ is a four dimensional complex matrix, and $a_{\alpha}^{\dagger}$ has four components, one on each leg ($r,u,l,d$).  
Due to the fermionic anti-commutation relations, we obtain that $T$ is an anti-symmetric matrix, reducing the number of allowed complex parameters to six. Solving the rotational invariance conditions of Eqs. (\ref{eq:Tpar}) and (\ref{eq:wpar}) reduces the number of free complex parameters in  $T$  to two:
\begin{equation}
    T = \mqty(
    0 & -z & -i y & -i z \\
    z & 0 & -i z & y \\
    i y & i z & 0 & z \\
    i z & -y & -z & 0 \\)  
    \label{eq:parametrization_minimal}
\end{equation}
and gives rise to 
\begin{equation}
    \begin{aligned}
        w\left(\mathbf{x},1\right) &= \exp\left(l^{\dagger}\left(\mathbf{x}+\hat{\mathbf{e}}_1\right)
        r^{\dagger}\left(\mathbf{x}\right)\right)\\
        w\left(\mathbf{x},2\right) &= \exp\left(\eta^2 
        d^{\dagger}\left(\mathbf{x}+\hat{\mathbf{e}}_2\right)u^{\dagger}\left(\mathbf{x}\right)\right).
    \end{aligned}
\end{equation}

It is easy to see that the strong coupling vacuum is included here, simply by choosing $y=z=0$, but which other states can be created with this choice? 
For that, we attempt to understand the meaning of the $y,z$ parameters. 
Recalling that the columns and rows of $T$ are ordered as $\left\{r,u,l,d\right\}$, we can see that the parameter $y$ relates to the creation of virtual fermions along a straight line ($r$ and $l$, or $u$ and $d$) while $z$ is associated with corners. 
To see this clearly, we can rewrite our $A\left(\mathbf{x}\right)$ operator as
\begin{widetext}
\begin{equation}
\begin{aligned}
A&=\left(1-2zr^{\dagger}u^{\dagger}\right)
\left(1-2iyr^{\dagger}l^{\dagger}\right)
\left(1-2izr^{\dagger}d^{\dagger}\right)
\left(1-2izu^{\dagger}l^{\dagger}\right)
\left(1+2yu^{\dagger}d^{\dagger}\right)
\left(1+2zl^{\dagger}d^{\dagger}\right)\\
&=\left(1-2z\begin{gathered}\includegraphics[scale=0.2]{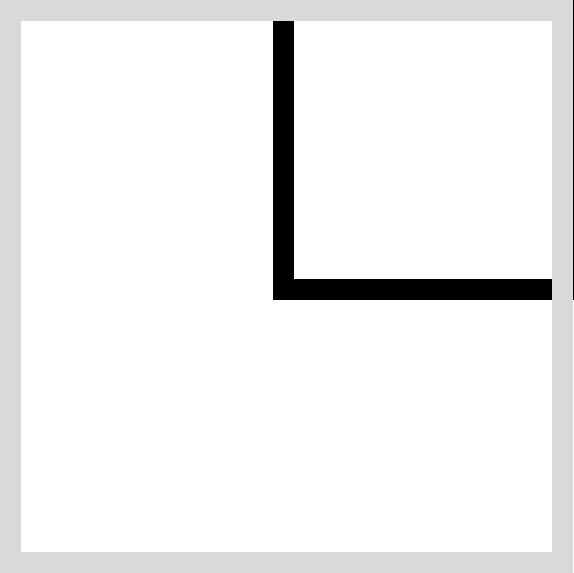}\end{gathered}\right)
\left(1-2iy\begin{gathered}\includegraphics[scale=0.2]{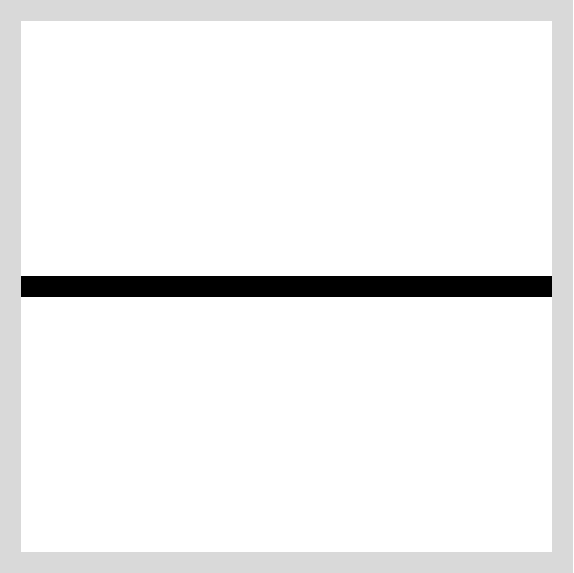}\end{gathered}\right)
\left(1-2iz\begin{gathered}\includegraphics[scale=0.2]{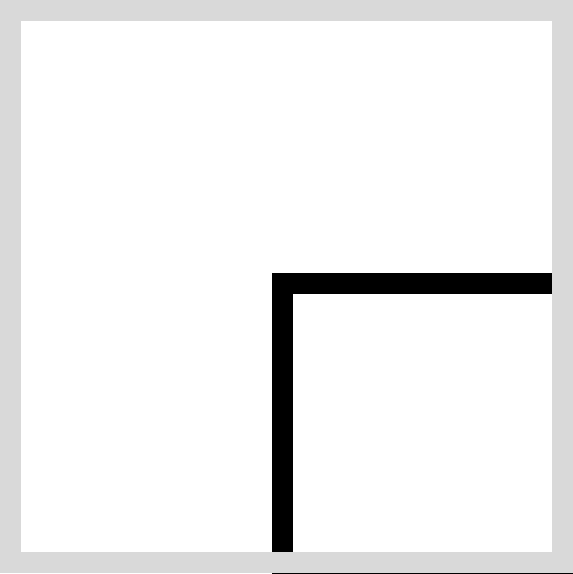}\end{gathered}\right)
\left(1-2iz\begin{gathered}\includegraphics[scale=0.2]{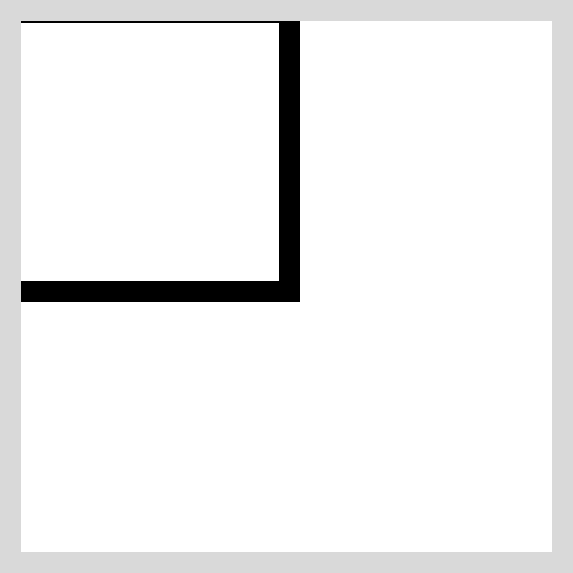}\end{gathered}\right)
\left(1+2y\begin{gathered}\includegraphics[scale=0.2]{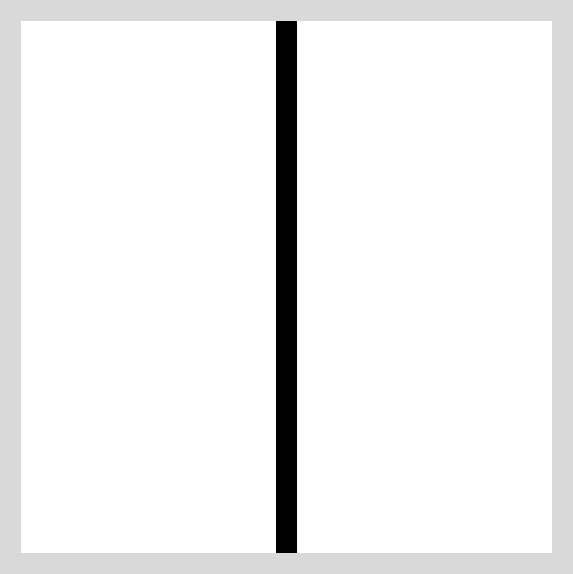}\end{gathered}\right)
\left(1+2z\begin{gathered}\includegraphics[scale=0.2]{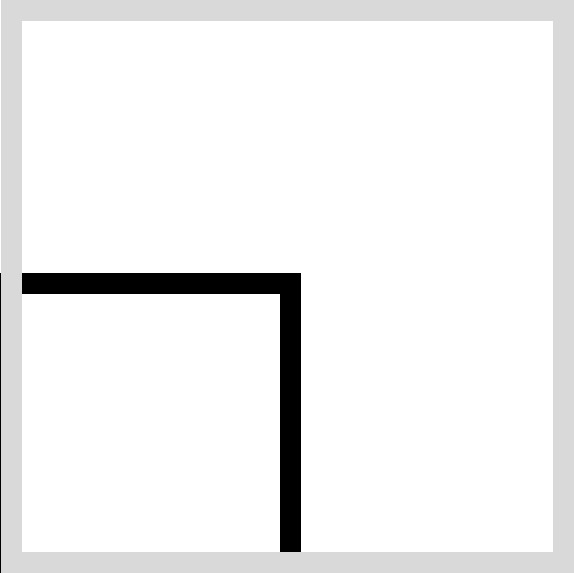}\end{gathered}\right)
\end{aligned}
\end{equation}
(we omitted the coordinate $\mathbf{x}$ which is shared by all operators for simplicity, and the graphical notation should be taken with caution, as it does not account for the ordering of creation operators - one can see the first row as defining proper ordering for the diagrams below).
We can open the brackets, and discover, due to the fermionic statistics and the fact the fermionic modes cannot be excited twice, that we have the following possibilities:
\begin{equation}
\begin{aligned}
A&=1-2zr^{\dagger}u^{\dagger}
-2iyr^{\dagger}l^{\dagger}
-2izr^{\dagger}d^{\dagger}
-2izu^{\dagger}l^{\dagger}
+2yu^{\dagger}d^{\dagger}
+2zl^{\dagger}d^{\dagger}
+4\left(iy^2-2z^2\right)r^{\dagger}u^{\dagger}l^{\dagger}d^{\dagger}
\\
&=1-2z\begin{gathered}\includegraphics[scale=0.2]{ru.png}\end{gathered}
-2iy\begin{gathered}\includegraphics[scale=0.2]{rl.png}\end{gathered}
-2iz\begin{gathered}\includegraphics[scale=0.2]{rd.png}\end{gathered}
-2iz\begin{gathered}\includegraphics[scale=0.2]{ul.png}\end{gathered}
+2y\begin{gathered}\includegraphics[scale=0.2]{ud.png}\end{gathered}
+2z\begin{gathered}\includegraphics[scale=0.2]{ld.png}\end{gathered}
+4\left(iy^2-2z^2\right)\begin{gathered}\includegraphics[scale=0.2]{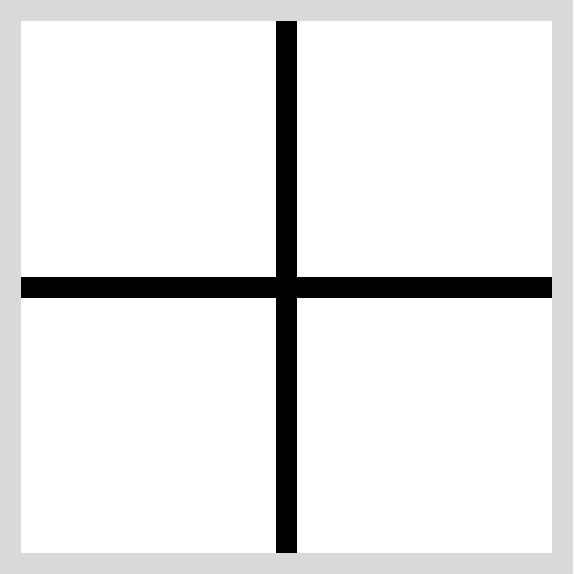}\end{gathered}.
\end{aligned}
\end{equation}
\end{widetext}

The PEPS $\ket{\psi}$ is created by a product of such operators on all sites, acting on the Fock vacuum and the strong coupling ground state $\ket{\psi_E}$. 
The gauging operator adds $Q$ on all the links where $r$ or $u$ are excited, and thus makes the ``physical flux map" look like the virtual one. 
The remaining action of $w^{\dagger}$ operators and projecting back on the virtual vaccum, completing the definition of the PEPS in Eq.~\eqref{eq:psidef}, guarantees gauge invariance and proper closing of the physical flux loops. 
We can use this to consider the structure of $\ket{\psi}$ as a series in the parameters $y,z$ (without assuming that any of them is small).

The zeroth order will be obtained by the $1$ ingredient of all $A$ and $w$ operators, and thus it will simply be 
\begin{equation}
    \ket{\psi^{(0)}}=\ket{\psi_E}.
\end{equation}
 
The next order will have to involve the shortest flux loop - a plaquette - requiring a contribution of four sites which is not $1$, and $1$ everywhere else. 
This will be a combination of four corners, and thus will be accompanied by $16z^4$. 
The final phase will be determined by the fermionic anticommutation rules when contracting with the right terms from $w^{\dagger}$; a straightfoward computation shows that
\begin{equation}
    \ket{\psi^{(1)}}=-16z^4\underset{p}{\sum}Q_{p_1}Q_{p_2}Q_{p_3}Q_{p_4}\ket{\psi_E}.
\end{equation}

We define the second order as that with two plaquettes excited. 
Here, there are several possibilities; first, of plaquettes $p,p'$ with no link or site in common. 
Then it is straightforward to show that the amplitude is the square of that for a single plaquette: $\left(-16z^4\right)^2 = 256z^8$. 
Another option is that of two plaquettes which share a link; then, we need four corners and two straight lines, and the amplitude per such a single excitement is $-64z^4y^2$; finally, we consider plaquettes with one site in common, which can be created in two ways, giving rise to two contributions to the amplitude: eight corners (contrbuting $256z^8$) or six corners and two straight lines ($-256i z^6y^2$). 

Continuing in this way is possible but tedious, so we stop here in order to evaluate what this state is able to describe. 
First, assume that $y=0$ and $\left|z\right|\ll 1$. 
Then the PEPS may be seen as a perturbative expansion, and we have
\begin{equation}
    \left|\psi\right\rangle = \left(1 - 16z^4\underset{p}{\sum}Q_{p_1} Q_{p_2} Q_{p_3}Q_{p_4}\right) \left|\psi_E\right\rangle + O\left(z^8\right).
\end{equation}
If we pick $z = \left(-128\lambda^2\right)^{-1/4} \ll 1$, we get, in leading order, the perturbative solution of the ground state for $\lambda \gg 1$ as in Eq.~\eqref{eq:strcoupl}. 
The next order, however, cannot be obtained, since we cannot excite two neighboring plaquettes with $y=0$, so we will not extend the discussion on this perturbative limit further.

Next, we consider the weak limit. 
We wish our ansatz to cover both extreme limits, so what about $\ket{\psi_B}$? 
We can take a look at its definition in Eq.~\eqref{eq:psiB} and start expanding all the brackets. 
We can clearly see that the superposition includes $\ket{\psi_E}$ as we have in our PEPS $\ket{\psi}$ and that all the orders (in particular the order of a single plaquette), no matter how many plaquettes are excited and where they are, have the same amplitude $1$. 
Going back to what we have just derived, this implies that we need $z^4=-\frac{1}{16}$. 
Going up to the second order, there are several kinds of terms; the pairs of far plaquttes carry an amplitude of $256z^8=1$ as required; but those who share a link carry an amplitude of $-64z^4y^2=-4y^2$, thus we require $y^2=-\frac{1}{4}$. 
However, we will run into contradiction with the other kind of terms, pairs of plaquettes with only a single site shared. 
Their amplitude will be $256\left(z^8-i z^6y^2\right)=1\pm 1 \neq 1$, and our conclusion is that the minimal ansatz does not cover the weak limit and thus does not satisfy our basic requirements from an ansatz.

\subsection{Optimized ansatz}

Next, we try to build an ansatz that will include $\ket{\psi_B}$, the weak limit ground state  too, as defined in~\eqref{eq:psiB} with two virtual fermions per leg ($F=2$). 
This time, we shall use a rather more constructive, bottom-up approach, by first considering a single plaquette.

\begin{figure}[t]
  \centering
  \includegraphics[width=0.6 \columnwidth]{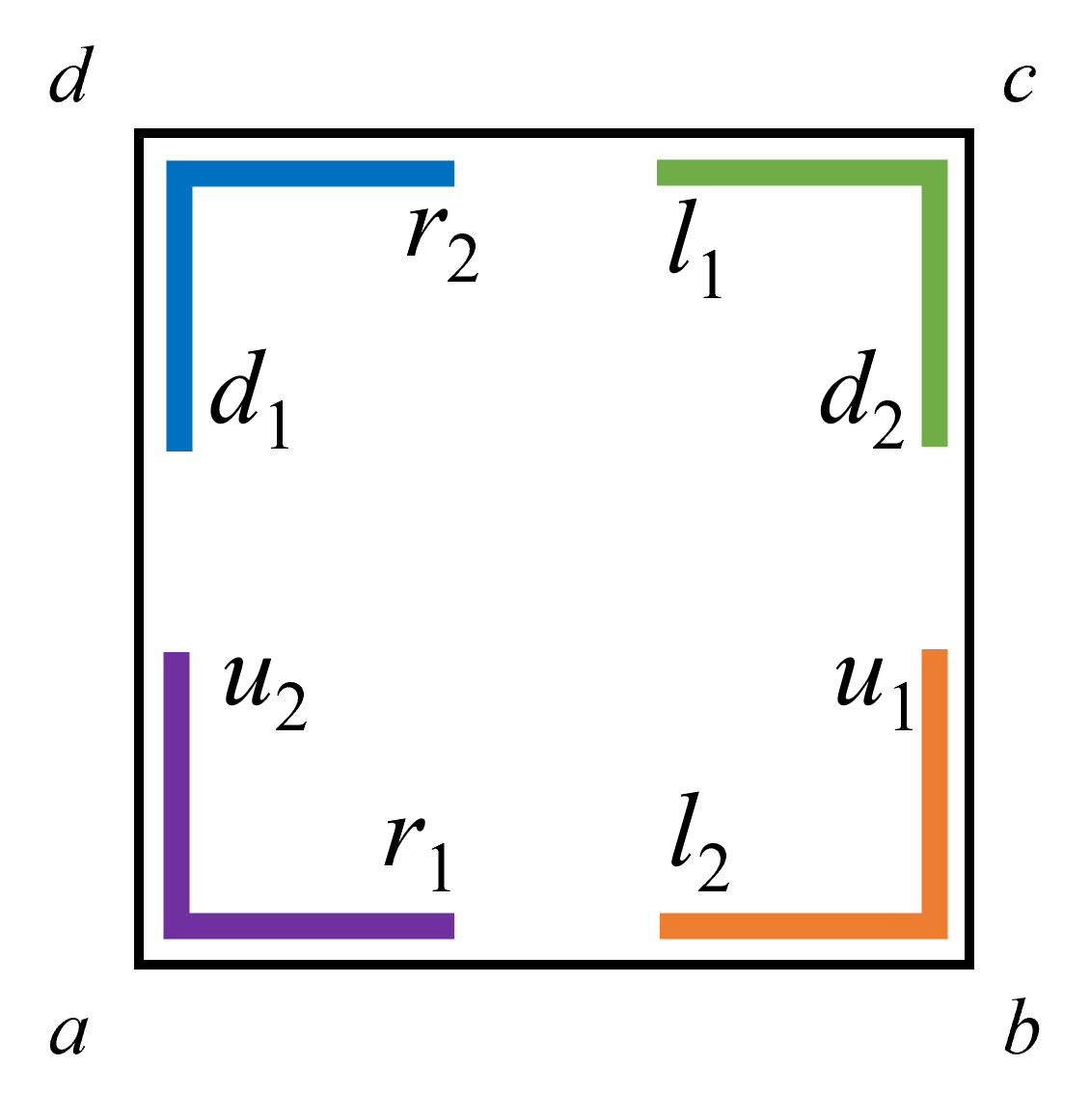}
  \caption{Notation convention for the single plaquette construction - the first step towards obtaining $\ket{\psi_B}$ with our ansatz.}
  \label{fig:plaqF2}
\end{figure}

Consider one plaquette and label the sites around it by $a,b,c,d$, as in Fig.~\ref{fig:plaqF2}. 
On each leg of the links we introduce a single virtual fermionic mode, eight altogether, and we name them as in the figure. 
Note that the sites $a,b,c,d$ are entirely independent of the parameters $a,b,c,d$ in Eq.~\eqref{eq:tmat_rotation_approach_ordered}.

Suppose this is our entire system, and we construct our PEPS for it; the initial gauge field state will simply be $\left|0\right\rangle = \left|p=0\right\rangle^{\otimes 4}$. 
The $A$ operators will take the most general forms
\begin{equation}
    \begin{aligned}
        A_{a} &= 1+f_{a} r\dgr_1\left(a\right)u\dgr_2\left(a\right) \\
        A_{b} &= 1+f_{b} u\dgr_1\left(b\right)l\dgr_2\left(b\right) \\
        A_{c} &= 1+f_{c} l\dgr_1\left(c\right)d\dgr_2\left(c\right) \\
        A_{d} &= 1+f_{d} d\dgr_1\left(d\right)r\dgr_2\left(d\right).
    \end{aligned}
\end{equation}
Gauging will be done as usual. In order to connect the modes properly, and following the previous construction, we define the operators
\begin{equation}
    \begin{aligned}
        w_{ab}&=\exp\left(l^{\dagger}_2\left(b\right)r^{\dagger}_1\left(a\right)\right)\\
        w_{bc}&=\exp\left(\eta^2d^{\dagger}_2\left(c\right)r^{\dagger}_1\left(b\right)\right)\\
        w_{dc}&=\exp\left(l^{\dagger}_1\left(c\right)r^{\dagger}_2\left(d\right)\right)\\
        w_{ad}&=\exp\left(\eta^2d^{\dagger}_1\left(d\right)u^{\dagger}_2\left(a\right)\right).
    \end{aligned}
\end{equation}

Finally, the PEPS for a single plaquette will take the form
\begin{equation}
    \ket{\psi_{\square}}=\bra{\Omega}
    w^{\dagger}_{ab}
    w^{\dagger}_{bc}
    w^{\dagger}_{dc}
    w^{\dagger}_{ad}
    \mathcal{U}_G
    A_{a}
    A_{b}
    A_{c}
    A_{d}
    \ket{\Omega}\ket{0}.
\end{equation}
It is easy to verify that the contraction of the virtual fermions will give rise to
\begin{equation}
    \ket{\psi_{\square}}=\left(1+
    f_{a}f_{b}f_{c}f_{d}
    Q_{ab}Q_{bc}Q_{dc}Q_{ad}\right)\ket{0}.
\end{equation}

Why did we bother to do all that? 
Because we wish to generate an operator which is a product of such operators on all the plaquettes,
\begin{equation}
    \mathcal{O}=\underset{p}{\prod}\left(1+f_{p_1}f_{p_2}f_{p_3}f_{p_4}
    Q_{p_1}Q_{p_2}Q_{p_3}Q_{p_4}\right);
\end{equation}
Clearly, when $f_{p_1}f_{p_2}f_{p_3}f_{p_4}=1$ for all the plaquettes, $\mathcal{O}\ket{\psi_E}=\ket{\psi_B}$.

In order to build this $\mathcal{O}$, we consider a PEPS $\ket{\psi}$ with $F=2$ - two virtual fermions per mode. 
On each site $\mathbf{x}$ we set
\begin{equation}
    A\left(\mathbf{x}\right) = 
    A_{a}\left(\mathbf{x}\right)
    A_{b}\left(\mathbf{x}\right)
    A_{c}\left(\mathbf{x}\right)
    A_{d}\left(\mathbf{x}\right)
\end{equation}
which allows a site to connect to all the four plaquettes around it - to one plaquette it plays the role of $a$ from the above construction, and to the others, the role of $b,c,d$. 
Finally, we introduce
\begin{equation}
    w\left(\mathbf{x},1\right)=\exp\left(l_2\left(\mathbf{x}+\hat{\mathbf{e}}_1\right)r_1\left(\mathbf{x}\right)\right)\exp\left(l_1\left(\mathbf{x}+\hat{\mathbf{e}}_1\right)r_2\left(\mathbf{x}\right)\right).
    \label{eq:w21}
\end{equation}
which accounts for the link being $ab$ for the plaquette on top of it, and $dc$ for the one beneath it. 
Similarly, 
\begin{equation}
    w\left(\mathbf{x},2\right)=\exp\left(\eta^2d_2\left(\mathbf{x}+\hat{\mathbf{e}}_2\right)u_1\left(\mathbf{x}\right)\right)\exp\left(\eta^2d_1\left(\mathbf{x}+\hat{\mathbf{e}}_2\right)u_2\left(\mathbf{x}\right)\right)
    \label{eq:w22}
\end{equation}
Plugging these into the PEPS construction $\ket{\psi}$ gives us $\ket{\psi_B}$ if we choose $f_{a}f_{b}f_{c}f_{d}=1$.

All we have to do is to show that this is embedded in some general parametrization with $F=2$. 
Such a construction will have an eight dimensional $T$ matrix. 
Demanding the rotation invariance condition~\eqref{eq:Tpar}, we obtain the $T$ matrix introduced in Eq.~\eqref{eq:tmat_rotation_approach_ordered}.
For the $w$ operators we stick to the choices of Eqs.~\eqref{eq:w21} and~\eqref{eq:w22}. 
They satisfy the rotation invariance conditions of Eq.~\eqref{eq:wpar}. 
Then, if we set all the parameters of the $T$ matrix (\ref{eq:tmat_rotation_approach_ordered}) to zero, other than $b$, we can get $\ket{\psi_B}$; inspecting Eq.~\ref{eq:tmat_rotation_approach_ordered} carefully and comparing it with our $\ket{\psi_B}$ construction implies that
\begin{equation}
    \begin{aligned}
        f_{a}&=2ib\\
        f_{b}&=-2b\\
        f_{c}&=-2ib\\
        f_{d}&=2b.
    \end{aligned}
\end{equation}
Taking the product, we obtain $f_{a}f_{b}f_{c}f_{d}=-16b^4$ and hence setting $b^4=-\frac{1}{16}$ and all the other parameters of~\eqref{eq:tmat_rotation_approach_ordered} to zero, gives us $\ket{\psi_B}$. Trivially, setting all the parameters to zero gives rise to $\ket{\psi_E}$.

Therefore, we choose the $F=2$ PEPS with $T$ given in~\eqref{eq:tmat_rotation_approach_ordered} and $w$ from~\eqref{eq:w21} and~\eqref{eq:w22}, or equivalently  \eqref{eq:Wmat}, as our ansatz. 
It satisfies all the requirements we set for ourselves: gauge invariant, rotation invariant, and including the extreme cases $\ket{\psi_E}$ and $\ket{\psi_B}$ with a minimal number of parameters. 

\begin{figure*}
    \centering
    \includegraphics[width=\textwidth]{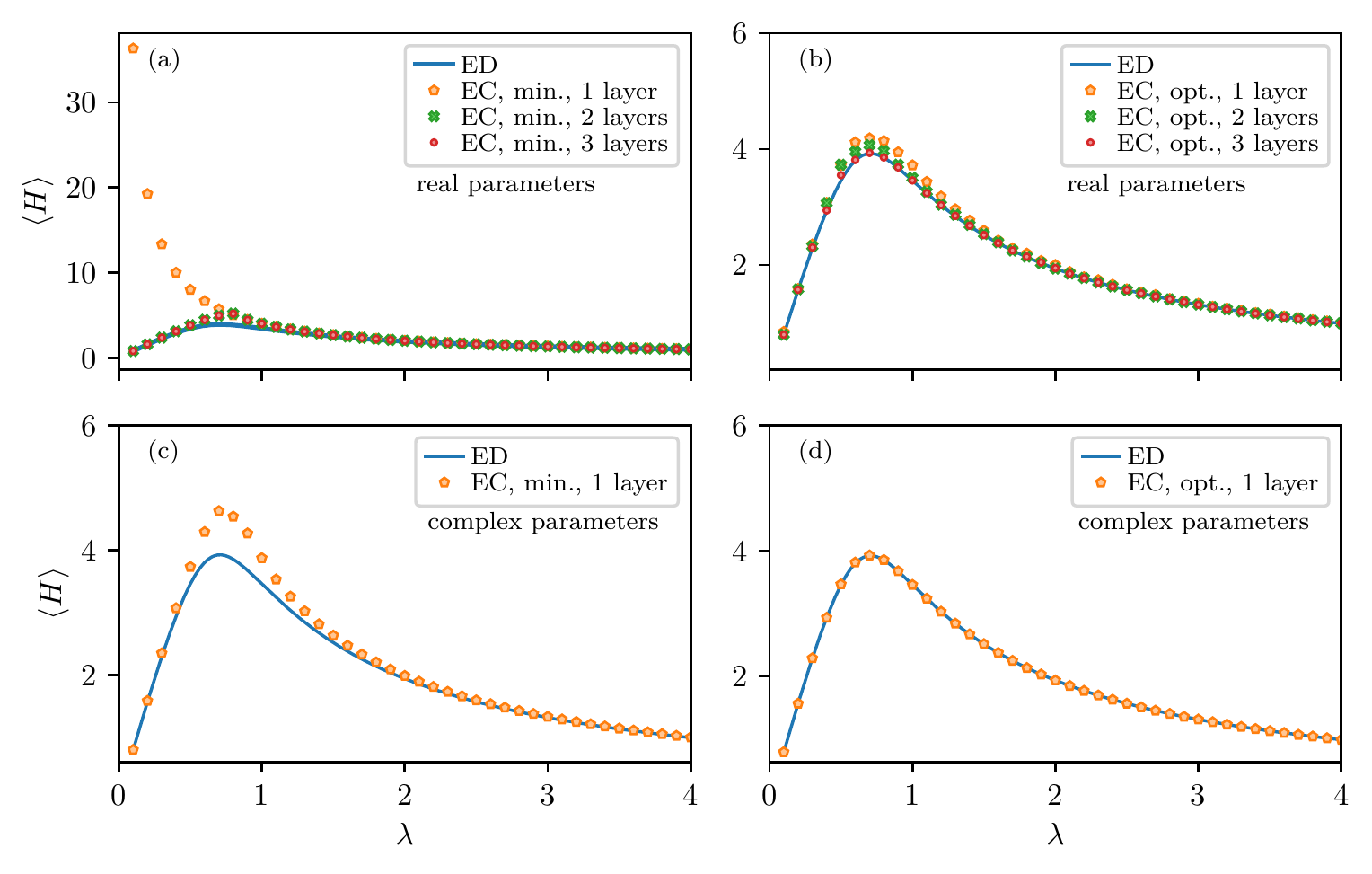}
    \caption{Comparison of different choices for the parameters and the number of virtual fermions on the link ($F$).
    The top(bottom) row shows the data for real(complex)-valued parameters.
    The left(right) column displays data for minimal(optimized) ansatz.
    All panels except for the top left one use the same scale for easier comparison.}
    \label{fig:real_vs_complex}
\end{figure*}

A big difference between the ansatz used in this work and previous work~\cite{zohar_fermionic_2015,emonts_variational_2020} is lifting the restriction to real parameters in $T$.
In Fig.~\ref{fig:real_vs_complex}, we illustrate the differences in using real/complex-valued parameters and changing the number of fermions on the links.
The mention of layers in the figure refers to the idea of enlarging the number of parameters by adding additional GGFPEPS coupled to the same gauge field.
In the absence of physical fermions the contraction of the GGFPEPS can be seen as a wave-function in the group element basis (cf. Eq.~\eqref{eq:psiQbasis}).
If we choose independent PEPS, we can compute all observables independently with only a linear runtime penalty in the number of layers.
For more details on layers, we refer to Ref.~\cite{emonts_variational_2020}.

In Fig.~\ref{fig:real_vs_complex}, Panel~(a) shows exact contraction data for the minimal ansatz with real parameters.
As in a previous work~\cite{emonts_variational_2020}, we see a $1/\lambda$ divergence for low couplings.
This is exactly the prefactor in front of $H_B$.
Thus, the ansatz state reaches a minimum energy for $H_B$ and fails to lower it further.
With an increase in the number of layers, the problem can be mitigated, but a distinct deviation exactly at the transition region remains.
Panel~(b) shows a similar plot for the optimized approach.
Even for the minimal number of a single layer, we get good agreement with the expected curve (solid line).
Upon increasing the number of layers, we see improvements in the transition region.
In panels~(c) and (d) the minimal and optimized approach are shown for complex parameters, respectively.
We note that the divergence at low couplings can be mitigated with just a single layer also in the $F=1$ case [panel (c)] for complex parameters.
The transition region, however, remains hard to access for the minimal approach.
Only the choice of complex parameters and the optimized ansatz of two virtual fermions on each link ($F=2$)[panel (d)] leads to a good match across the full range of couplings.

\section{Computation of the covariance matrix in Dirac modes\label{app:covariance_matrices}}
We consider a fermionic Gaussian state of the form
\begin{align}
  \ket{\psi}=\mathcal{N} \exp(\frac{1}{2}M_{ij}a_{i}\dgr a_{j}\dgr)\ket{\Omega}
\end{align}
where $\{a_i\dgr\}_{i=1}^{2R}$ create fermionic modes from the Fock vacuum $\ket{\Omega}$ and $M_{ij}=-M_{ji}$ is an anti-symmetric matrix.

The anti-symmetric matrix $M$ can always be brought to a canonical form $M_0$, with a unitary $U$
\begin{align}
  M=U\tran M_0 U.
  \label{eq:app_canoncial_m}
\end{align}
If $M_{ij}\in \mathbb{R}$, $U$ is orthogonal, but the decomposition in~\eqref{eq:app_canoncial_m} involves $U\tran$ and not $U\dgr$ for both $\mathbb{R}$ and $\mathbb{C}$.

In both cases ($M\in\mathbb{R}$ and $M\in\mathbb{C}$),
\begin{align}
  M_0=\mqty(0&\lambda_1&&&\\
  -\lambda_1&0&&&\\
  &&&\ddots&&\\
  &&&&0&\lambda_R\\
  &&&&-\lambda_R&0)
\end{align}
with $\lambda_k\in\mathbb{R}$ and $\lambda_k\leq 0$.
We denote $M_0$ in terms of a direct sum
\begin{align}
  M_0=\bigoplus_{k=1}^R M_0(k) 
\end{align}
where $M_0(k)=i\lambda_k \sigma_y$.

In this canonical basis, $\ket{\psi}$ can be written as a product of BCS states $\ket{\psi_k}$:
\begin{align}
  \ket{\psi}&=\bigotimes_{k=1}^{R}\ket{\psi_k}\qq{with}\\
  \ket{\psi_k}&=\frac{1}{\sqrt{1+\lambda_k ^2}}\left( 1+\lambda_k  b_{2k-1}\dgr b_{2k}\dgr \right) \ket{\Omega_k}
\end{align}
and $b_i\dgr=U_{ij}a_j\dgr$.

The covariance matrix of $\ket{\psi}$ in this canonical basis, $\Gamma_0$, can be written as a direct sum:
\begin{align}
  \Gamma_0=\bigoplus_{k=1}^{R}\Gamma_0(k)
\end{align}
where $\Gamma_0(k)$ is the $4\times 4$ covariance matrix of the BCS state $\ket{\psi_k}$.

The Dirac covariance matrix for $\ket{\psi_k}$ is given as
\begin{align}
  \Gamma_{\alpha\beta}^{D}=\mqty(Q_k&R_k\\\overline{R}_k&\overline{Q}_k)
\end{align}
where
\begin{align}
    \begin{aligned}
        Q_{\alpha\beta}(k)&=\frac{i}{2}\expval{\comm{\tilde{b}_\alpha(k)}{\tilde{b}_\beta(k)}}{\psi_k}\\
        R_{\alpha\beta}(k)&=\frac{i}{2}\expval{\comm{\tilde{b}_\alpha(k)}{\tilde{b}\dgr_\beta(k)}}{\psi_k}\\
        \tilde{b}_1(k)&=b_{2k-1}\\
        \tilde{b}_2(k)&=b_{2k}.
    \end{aligned}
\end{align}
For $\ket{\psi_k}=\left( u(k) + v(k) b_1\dgr(k) b_2\dgr(k) \right)$,
\begin{align}
    \begin{aligned}
        Q(k)&=u(k)v(k)\sigma_y\\
        &=-i(1-M_0^2(k))\inv M_0(k)
    \end{aligned}\\
    \begin{aligned}
        R(k)&=\frac{i}{2}\left( 1-2v^2(k) \right)\id\\
        &=\frac{i}{2}\left(1-M_0^2(k)\right)\inv \left( 1+M_0^2(k) \right)
    \end{aligned}
\end{align}
with $u(k)=\frac{1}{\sqrt{1+\lambda_k ^2}}$ and $v(k)=\frac{\lambda_k }{\sqrt{1+\lambda_k ^2}}$ and
$1-M_0^2(k)=\left(1+\lambda_k^2\right)\id$.

Using the structure of the direct sum for $R$ and $Q$, we obtain
\begin{align}
  R&=\frac{i}{2}(1-M_0^2)\inv(1+M_0^2)\\
  Q&=-i(1-M_0^2)\inv M_0.
\end{align}
In the original ordering of the operators $b$ and $b\dgr$ we get
\begin{align}
  \Gamma_0^D&=\mqty(Q&R\\\overline{R}&\overline{Q})\\
  &=i\mqty(-(1-M_0^2)\inv M_0& \frac{1}{2}(1-M_0^2)\inv (1+M_0^2)\\
  -\frac{1}{2}(1-\overline{M}_0^2)\inv (1+\overline{M}_0^2)&(1-\overline{M}_0^2)\inv \overline{M}_0).
\end{align}
Here, we ordered the operators such that $\{b_1,\cdots,b_R,b_1\dgr,\cdots,b_R\dgr\}$.
Although all $\lambda_k\in\mathbb{R}$, we keep the notation $\overline{M}_0$ for easier notation later.

We rotate back to the $a$ basis that we started with and define
\begin{align}
  V=\mqty(\dmat{\overline{U},U}),\, \vec{a}=\mqty(a_1\\\vdots\\a_{2R}\\a_1\dgr\\\vdots\\a_{2R}\dgr),\,\vec{b}=\mqty(b_1\\\vdots\\b_{2R}\\b_1\dgr\\\vdots\\b_{2R}\dgr)
\end{align}
such that
\begin{align*}
  \vec{b}&=V\vec{a}\\
  \vec{a}&=V\dgr\vec{b}=\mqty(\dmat{U\tran,U\dgr})\vec{b}\dgr.
\end{align*}
In the $a$ basis, we obtain
\begin{align}
    \begin{aligned}
        \Gamma_{\alpha\beta}^D&=\frac{i}{2}\expval{\comm{\vec{a}_\alpha}{\vec{a}_\beta}}{\psi}\\
        &=\frac{i}{2}\expval{\comm{V_{\alpha\alpha'}\vec{b}_{\alpha'}}{V_{\beta\beta'}}\vec{b_{\beta'}}}{\psi}\\
        &=(V\dgr \Gamma_0^D \overline{V})_{\alpha\beta}
    \end{aligned}
\end{align}
which evaluates to
\begin{align}
    \Gamma^D=i\mqty(-M^- M& \frac{1}{2}M^- \left(1+M\overline{M}\right)\\
    -\frac{1}{2}M^- \left(1+\overline{M}M\right) & M^- \overline{M}).
    \label{eq:app_gamma_matrix_in_t}
\end{align}
with $M^-=\left(1-M\overline{M}\right)\inv$.

The covariance matrix in terms of Majorana modes can be computed by a linear transformation from Eq.~\eqref{eq:app_gamma_matrix_in_t} which follows directly from the definition of the Majorana modes in Eq.~\eqref{eq:app_def_majorana_modes}.

The direct connection between the parametrization $M$ and the Majorana covariance matrix enables us to compute the derivatives of the covariance matrices automatically by symbolic differentiation.
Since the matrices are not modified during a Monte Carlo evaluation, we can store the numerical evaluation of the derivatives for a given set of parameters.

\section{Calculation of $\expval{P}$\label{app:electric_energy}}
Here we shall elaborate and give detail on the computation of $F_P\left(\mathcal{Q}\right)$, that we need for the computation of $\expval{P}$ and hence for the electric energy part of $\expval{H}$. 
Following Eq.~\eqref{eq:Pexp}, recall that
\begin{equation}
    F_P\left(\mathcal{Q}\right) 
    =\frac{
    \overline{\psi\left(\hat{Q},q-1\right)}
    \psi\left(\hat{Q},q\right)}
    {\abs{\psi\left(\mathcal{Q}\right)}^2}.
\end{equation}

If we define
\begin{equation}
\ket{\phi\left(\mathbf{\mathcal{Q}}\right)} = \mathcal{U}_{\mathcal{Q}} \underset{\mathbf{x}}{\prod}A\left(\mathbf{x}\right)\ket{\Omega},
\end{equation}
$\Omega\left(\mathbf{x},i\right)$ as the projector onto the empty state of all the virtual fermionic modes on the link $\left(\mathbf{x},i\right)$,
and
\begin{equation}
\omega\left(\mathbf{x},i\right)=w\left(\mathbf{x},i\right)\Omega\left(\mathbf{x},i\right)w^{\dagger}\left(\mathbf{x},i\right),
\end{equation}
we can express the numerator of $F_P\left(\mathcal{Q}\right)$ as
\begin{align}
    \begin{aligned}
        \overline{\psi\left(\hat{Q},q-1\right)} \psi\left(\hat{Q},q\right)=&
        \bra{\phi\left(\mathbf{\mathcal{Q}}\right)}
        \exp\left(\overset{F}{\underset{\alpha=1}{\sum}}i\pi r\dgr_{\alpha} r_{\alpha} \right)\times\\
        &\underset{\vb{x},i}{\prod} \omega\left(\vb{x},i\right)
        \ket{\phi\left(\mathbf{\mathcal{Q}}\right)} 
    \end{aligned}
\end{align}
(without loss of generality - due to the rotational symmetry - we assume that we compute the electric energy for a horizontal link).
The numerator is the unnormalized expectation value of the Gaussian operator 
$\exp\left(\overset{F}{\underset{\alpha=1}{\sum}}i\pi r\dgr_{\alpha} r_{\alpha} \right) \underset{\mathbf{x},i}{\prod} \omega\left(\mathbf{x},i\right)$
with respect to the state $\ket{\phi\left(\gauge\right)}$.

For simplicity, let us first assume that $F=1$ - the minimal ansatz introduced above.
If we denote by $\tilde{\omega}$ the product of $\omega\left(\mathbf{x},i\right)$ on all the links but the one where the expectation value is computed, we obtain for the numerator
\begin{align}
    \begin{aligned}
        &\overline{\psi\left(\hat{Q},q-1\right)} \psi\left(\hat{Q},q\right)\\
        =&\frac{1}{2} \bra{\phi\left(\gauge\right)}
        \left(rl-l\dgr r\dgr-1+ll\dgr+rr\dgr\right)\tilde{\omega}
        \ket{\phi\left(\gauge\right)}\\
        \propto&\frac{1}{2} \bra{\phi\left(\gauge\right)}
        \tilde{\omega}\left(rl-l\dgr r\dgr-1+ll\dgr+rr\dgr\right)\tilde{\omega}
        \ket{\phi\left(\gauge\right)} \\
        =&\frac{1}{2} \bra{\tilde{\phi}}
        \left(rl-l\dgr r\dgr-1+ll\dgr+rr\dgr\right)
        \ket{\tilde{\phi}} 
    \end{aligned}
    \label{eq:app_el_energy_dirac_minimal}
\end{align}
where $r,l$ are the annihilation operators of the virtual fermions on both sides of the link we study (belonging to two neighboring sites).
In the third line of~\eqref{eq:app_el_energy_dirac_minimal}, we used the fact that $\tilde{\omega}^2\propto\tilde{\omega}$ (as a non-normalized density matrix of a pure state), as well as that $\tilde{\omega}$ commutes with the modes in parentheses because it acts on different links.
For convenience, we define the state \begin{equation} 
\ket{\tilde{\phi}}=\tilde{\omega}\ket{\phi(\gauge)}.
\end{equation}
The quantity we wish to compute is the expectation value of an operator acting on the $l,r$ modes of one particular link, with respect to this state. 
Since  $\ket{\phi(\gauge)}$ is Gaussian, this can be extracted from its covariance matrix.

All covariance matrices so far have been formulated in terms of Majorana modes.
Thus, we express the operator whose expectation value we seek in~\eqref{eq:app_el_energy_dirac_minimal} in terms of Majorana modes: 
\begin{align}
    \begin{aligned}
        &rl-l\dgr r\dgr - 1+ll\dgr+rr\dgr\\
        =&\frac{1}{2}(\maj{r}{1}\maj{l}{1} - \maj{r}{2}\maj{l}{2} + i\maj{l}{1}\maj{l}{2} + i\maj{r}{1}\maj{r}{2})\, .
    \end{aligned}
    \label{eq:app_el_energy_maj}
\end{align}

In a previous work~\cite{emonts_variational_2020}, we rewrote the operator in \eqref{eq:app_el_energy_maj} in terms of Grassman variables and obtained a new form for the gauged covariance matrix of the projectors $\gammain$.
The resulting expression required the computation of a system-sized Pfaffian which is computationally expensive since it must be recomputed with every measurement.
To our knowledge, there is no algorithm to infer the new value of a Pfaffian after a local change in a matrix~\cite{wimmer_algorithm_2012}.
In the following, we present a different method that  explicitly uses the Gaussian character of the GGFPEPS.

\begin{figure}[h]
  \centering
  \includegraphics[width=0.6 \columnwidth]{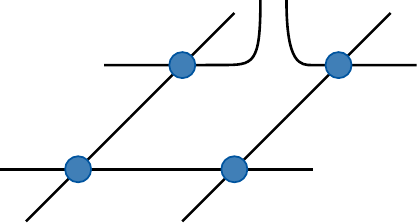}
  \caption{
      Rearrangement of the contraction pattern in terms of PEPS contractions. 
      The link at the top is selected for the electric energy computation.
      Its legs are kept open to use the Gaussian mapping.
  }
  \label{fig:app_peps_contraction_el_energy}
\end{figure}

Using standard fermionic Gaussian states \cite{bravyi_lagrangian_2005}, and in particular fermionic Gaussian PEPS, techniques \cite{kraus_fermionic_2010}, we can compute the covariance matrix of $\ket{\phi(\gauge)}$ using the so-called Gaussian map; such a Gaussian map takes as its input the state $\tilde{\omega}$, involving the modes on all the links but the one we are interested in, and it is parameterized by the covariance matrix $D$ of the state $\ket{\phi(\gauge)}$. 
We sort the fermionic modes into two groups: the ones which are contracted - those on all the links but the one we look at, and the non-contracted, or open ones, which are all the rest (mathematically, this is equivalent to virtual and physical modes respectively, in conventional fermionic PEPS constructions \cite{kraus_fermionic_2010}) - see Fig.~\ref{fig:app_peps_contraction_el_energy} for  graphical explanation.

\begin{figure}[h]
  \centering
  \subfloat{\includegraphics[width=0.45\columnwidth]{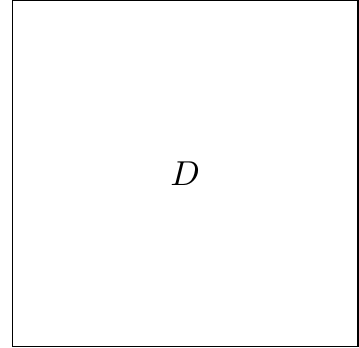}}
  \hfill
  \subfloat{\includegraphics[width=0.45\columnwidth]{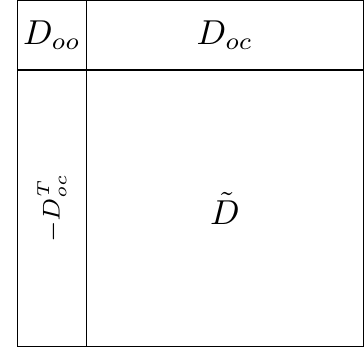}}
  \caption{Left: Covariance matrix in the standard case. Right: Adapted covariance matrix to trace out the environment of a single link.
  The subscripts $o$ and $c$ refer to open and contracted modes, respectively.}
  \label{fig:covariance_matrices}
\end{figure}

The output of the Gaussian map is the covariance matrix of $\ket{\tilde{\phi}}$, which belongs to the Hilbert space of the non-contracted fermions \footnote{Strictly speaking, the contracted fermions are there too but they are in a trivial product state with the rest and can be ignored, as can be read directly from \ref{eq:app_el_energy_dirac_minimal}.}. It is given by \cite{bravyi_lagrangian_2005,kraus_fermionic_2010}
\begin{align}
    \Gamma_v&= D_{oo} + D_{oc}\left( \tilde{D} - \gammainmod\left(
    \mathcal{Q}
    \right)\right)\inv D_{oc}\tran ,
\end{align}
where $D_{oo}$, $D_{oc}$ and $\tilde{D}$ are the blocks of the covariance $D$ containing correlations of open modes with themselves, open with contracted, and contracted modes with themselves, respectively, as depicted in Fig.~\ref{fig:covariance_matrices}. 
$ \gammainmod\left(\mathcal{Q} \right)$ is the covariance matrix of $\tilde{\omega}$.

Using the basis $\left\{\maj{l}{1},\maj{l}{2},\maj{r}{1},\maj{r}{2}\right\}$ for the modes, we can write the matrix $\Gamma_v$ as 
\begin{align}
  \Gamma_v=\frac{1}{\braket{\tilde{\phi}}}\left(
  \begin{array}{cccc}
        0            & i\expval{\maj{l}{1}\maj{l}{2}}     & i\expval{\maj{l}{1}\maj{r}{1}} & i\expval{\maj{l}{1}\maj{r}{2}}     \\
                     &          0                              & i\expval{\maj{l}{2}\maj{r}{1}} & i\expval{\maj{l}{2}\maj{r}{2}}     \\
                     &                                         &       0                             & i\expval{\maj{r}{1}\maj{r}{2}}     \\
                     &                                         &                                     &         0
  \end{array}
  \right)\, ,
\end{align}
with $\expval{\cdot}=\expval{\cdot}{\tilde{\phi}}$.
Since $\Gamma_v$ is anti-symmetric, we do not fill out the lower half of the matrix.

We identify the expression~\eqref{eq:app_el_energy_maj} in terms of elements of the covariance matrix
\begin{align}
    \begin{aligned}
      &\bra{\tilde{\phi}}\frac{1}{2}\left( \maj{r}{1}\maj{l}{1}-\maj{r}{2}\maj{l}{2}+i\maj{l}{1}\maj{l}{2} + i\maj{r}{1}\maj{r}{2} \right)\ket{\tilde{\phi}} \\
      =& \frac{1}{2}\left( -\frac{1}{i}\Gamma_{v;1,3} +\frac{1}{i}\Gamma_{v;1,4} +\Gamma_{v;1,2} +\Gamma_{v;3,4} \right)\braket{\tilde{\phi}}\\
      =&\frac{1}{2}\left( i\Gamma_{v;1,3} - i\Gamma_{v;1,4} + \Gamma_{v;1,2} +\Gamma_{v;3,4} \right)\braket{\tilde{\phi}}\, .
  \end{aligned}
\end{align}

The full expression of $\expval{P}$ reads
\begin{align}
  \expval{P}\nonumber &=\frac{\bra{\psi}P\ket{\psi}}{\braket{\psi}}\nonumber\\
  &=\sum_{q,\hat{Q}}\frac{1}{4}\frac{\left( i\Gamma_{v;1,3} - i\Gamma_{v;1,4} + \Gamma_{v;1,2} +\Gamma_{v;3,4} \right)\braket{\tilde{\phi}}}{\bra{\psi(\hat{Q},q)}\ket{\psi(\hat{Q},q)}}p(\gauge)\, .
\end{align}

Finally, we can directly remove the anti-hermitian part since $P$ is a Hermitian operator in $\Z{2}$ and it will add up to zero in the end. Thus
\begin{align}
  \expval{P}\nonumber &=\frac{\bra{\psi}P\ket{\psi}}{\braket{\psi}}\nonumber\\
  &=\sum_{q,\hat{Q}}\frac{1}{4}\frac{\left( \Gamma_{v;1,2} +\Gamma_{v;3,4} \right)\braket{\tilde{\phi}}}{\bra{\psi(\hat{Q},q)}\ket{\psi(\hat{Q},q)}}p(\gauge)\, .
  \label{eq:app_el_energy_op_1_minimal}
\end{align}
Here, the norm $\braket{\phi}$ can be computed using~\eqref{eq:overlap_default} for the modified matrices $\tilde{D}$ and $\gammainmod$.

In the case of the optimized approach with $F=2$, the  expression corresponding to~\eqref{eq:app_el_energy_maj} is
\begin{align}
    \begin{aligned}
        &e^{i\Phi r_1\dgr r_1}e^{i\Phi r_2\dgr r_2} \frac{1}{4} \left(1+l_1\dgr r_2\dgr\right)\left(1+l_2\dgr r_1\dgr\right) \times\\
        &r_1 r_1\dgr r_2r_2\dgr l_1l_1\dgr l_2l_2\dgr \left(1+r_2l_1\right)\left(1+r_1l_2\right)\\
        =&\frac{1}{4} \left(\maj{r}{1}_1 \maj{l}{1}_2 - \maj{r}{2}_1 \maj{l}{2}_2 + i \maj{l}{1}_2 \maj{l}{2}_2 + i \maj{r}{1}_1 \maj{r}{2}_1\right)\times\\
        &\frac{1}{4} \left(\maj{r}{1}_2 \maj{l}{1}_1 - \maj{r}{2}_2 \maj{l}{2}_1 + i \maj{l}{1}_1 \maj{l}{2}_1 + i \maj{r}{1}_2 \maj{r}{2}_2\right)\, .
        \end{aligned}
    \label{eq:el_energy_maj_2cp}
\end{align}
The subscript indices denote the different values of $F$ in the system.

While we can write the expectation value in terms of a multiplication Majorana modes, we cannot directly transform the products into elements of the covariance matrix.
The products of four Majorana modes are identified with Pfaffians in terms of submatrices of the covariance matrix $D$ using equation~(17) from~\cite{bravyi_lagrangian_2005}

The equation~\cite{bravyi_lagrangian_2005}
\begin{align}
  \tr(\rho i^p c_{a_1}c_{a_2}\cdots c_{2p})=\Pf(\Gamma_v\vert_{a_1,\cdots,a_{2p}})\, ,
  \label{eq:pfaffian_submatrices}
\end{align}
with $1\leq a_1 < \cdots < a_{2p}\leq 2n$, enables the transformation from products of Majorana modes to matrix elements of the covariance matrix.
Here, $\Gamma_v\vert_{a_1,\cdots,a_{2p}}$ is the $2p\times 2p$ submatrix with the indicated rows and columns. 
More concretely, we need the contraction of $4$ Majorana modes
\begin{align}
  \Tr(\rho c_1 c_2 c_3 c_4)=-\Pf(\Gamma_v\vert_{1234})\, .
\end{align}

In terms of Pfaffians, we can write Eq.~\eqref{eq:el_energy_maj_2cp} as
\begin{widetext}
\begin{align}
    \begin{aligned}
      &\frac{1}{16} 
      [\Pf(\Gamma_v\vert_{1357}) - \Pf(\Gamma_v\vert_{2358}) - i \Pf(\Gamma_v\vert_{1235}) - i \Pf(\Gamma_v\vert_{3578})\\
       &- \Pf(\Gamma_v\vert_{1467}) +  \Pf(\Gamma_v\vert_{2468})+ i  \Pf(\Gamma_v\vert_{1246}) + i \Pf(\Gamma_v\vert_{4678})\\
       &+i \Pf(\Gamma_v\vert_{1567}) - i \Pf(\Gamma_v\vert_{2568}) + \Pf(\Gamma_v\vert_{1256}) +   \Pf(\Gamma_v\vert_{5678})\\
       &+i \Pf(\Gamma_v\vert_{1347}) - i \Pf(\Gamma_v\vert_{2348}) + \Pf(\Gamma_v\vert_{1234}) +   \Pf(\Gamma_v\vert_{3478})]\, .
   \end{aligned}
   \label{eq:app_pfaffians_optimized}
\end{align}
The final expression of $\expval{P}$ reads, after omitting all imaginary terms, 
\begin{align}
    \begin{aligned}
      \expval{P}\nonumber &=\frac{\bra{\psi}P\ket{\psi}}{\braket{\psi}}\nonumber\\
      &=\frac{1}{16}\sum_{q,\hat{Q}}\left[\Pf(\Gamma_v\vert_{1357}) - \Pf(\Gamma_v\vert_{2358})- \Pf(\Gamma_v\vert_{1467}) +  \Pf(\Gamma_v\vert_{2468})\right.\\
      &\phantom{=\frac{1}{16}\sum_{q,\hat{Q}}}\left.+\Pf(\Gamma_v\vert_{1256}) +   \Pf(\Gamma_v\vert_{5678})+ \Pf(\Gamma_v\vert_{1234}) -   \Pf(\Gamma_v\vert_{3478})\right]\frac{\braket{\tilde{\phi}}}{\braket{\psi(\hat{Q},q)}}p(\hat{Q},q)\, .
  \end{aligned}
\end{align}
\end{widetext}

\section{Calculation of derivatives\label{app:derivative}}

The idea of a variational algorithm is to minimize the energy by (iteratively) adapting the parameters of a state.
While gradient free optimizations are possible, the gradient of the energy with respect to the parameters, speeds up the minimization significantly.
The total energy of the system consists of two parts: $H=H_E+H_B$.
Since the electric energy is non-diagonal in the group element basis (cf. sec.~\ref{sec:p_exp_value}), we go through the relevant calculations in detail.
The easier case of the magnetic energy (diagonal in the group element basis) follows directly.

We consider an observable $O$ and its expectation value
\begin{align}
  \expval{O}=\sum_{\gauge} F_{O}(\gauge)p(\gauge).
\end{align}
Our aim is to calculate the derivative with respect to a parameter $\alpha$ of the matrix $T(\alpha)$.
The number of parameters depends on the value of $F$.
The derivative of the observable can be written as 
\begin{align}
    \begin{aligned}
        \pdv{\alpha}\expval{O}=&\expval{\pdv{\alpha}O}+
        \expval{O\frac{\pdv{\alpha}\normsq}{\normsq}}\\
        &-\expval{O}\expval{\frac{\pdv{\alpha}\normsq}{\normsq}}\, .
    \end{aligned}
    \label{eq:app_observable_derivative}
\end{align}
The expression can be derived by considering the logarithmic derivative  $\pdv{\alpha}\ln\expval{O}$ and transforming back at the end of the calculation.
If the observable does not explicitly depend on the parameters, as is the case for Wilson loops, the first term in~\eqref{eq:app_observable_derivative} vanishes.

In a first step, we focus on the calculation of the second and third term which are always present since the norm always depends on the parameters.
Instead of directly tracking the derivative of the parameters through the state construction, we will use the chain rule.
The matrix $D$ does not change during one Monte Carlo computation with a given set of parameters and is the only one that contains the parameters. 

The derivative of the norm is
\begin{align}
    \begin{aligned}
        &\pdv{\alpha}\normsq\\
        =&\dv{\alpha}\sqrt{\det(\frac{1-\gammain(\gauge) D(\alpha)}{2})} \\
        =&-\frac{1}{2^{N+1}} \sqrt{\det(1-\gammain D)} \times\\
        &\Tr(\gammain(\gauge) {\pdv{D}{\alpha}} (1-\gammain(\gauge) D)^{-1}) \, .
    \end{aligned}
\end{align}
The calculation is described in more detail in an Appendix of Ref.~\cite{emonts_variational_2020}.

According to the calculation in Appendix~\ref{app:electric_energy}, the electric energy is obtained by changing the Gaussian map.
The expression for the electric energy contains the covariance matrix and two norms $\braket{\psi}$ and $\braket{\tilde{\phi}}$ that depend on the parameters.
The exact expression depends on the ansatz. 

For the minimal ansatz ($F=1$), we compute the observable of the electric energy with [cf. Equation~\eqref{eq:app_el_energy_maj}]
\begin{align}
  F_{P}=\frac{\frac{1}{4}\left(\Gamma_{v;1,2} +\Gamma_{v;3,4} \right)\braket{\phi}}{\braket{\Psi(\gauge,q)}}\, .
\end{align}
The derivative of the expression reads
\begin{align}
    \begin{aligned}
        \pdv{F_{P}}{\alpha}&=\frac{1}{4}\pdv{\alpha}\frac{\left( \Gamma_{v;1,2} + \Gamma_{v;3,4} \right)\braket{\phi}}{\braket{\psi}}\\
        &=\frac{1}{4}\frac{\pdv{\alpha}\left[ \Gamma_{v;1,2} + \Gamma_{v;3,4} \right]\braket{\phi}}{\braket{\psi}}+F_{P}\left( \tilde{v}-v \right)
    \end{aligned}
    \label{eq:app_deriv_f_el_minimal}
\end{align}
with $v$ and $\tilde{v}$ given by 
\begin{align}
    \begin{aligned}
        v&=-\frac{1}{2}\Tr\left(\gammain\pdv{D}{\alpha}D^{-1}\left(D^{-1}-\gammain\right)^{-1}\right)\\
        \tilde{v}&=-\frac{1}{2}\Tr\left(\gammainmod\pdv{\tilde{D}}{\alpha}\tilde{D}^{-1}\left(\tilde{D}^{-1}-\gammainmod\right)^{-1}\right)
    \end{aligned}
\end{align}
The tilde decorations used here are those introduced in Appendix \ref{app:electric_energy}.

When using the optimized ansatz ($F=2$), we have a slightly different expression for the electric energy
\begin{align}
  F_{P}=\frac{\frac{1}{16}f(\Gamma_v)\braket{\phi}}{\braket{\Psi(\gauge,q)}}
\end{align}
where $f(\Gamma_v)$ is the sum of Pfaffians in~\eqref{eq:app_pfaffians_optimized} without the prefactor of $\frac{1}{16}$.

The derivative of the expression is structurally similar to one of the minimal approach
\begin{align}
    \begin{aligned}
        \pdv{F_{P}}{\alpha}&=\frac{1}{16}\pdv{\alpha}\frac{f(\Gamma_v) \braket{\phi}}{\braket{\psi}}\\
        &=\frac{1}{16}\frac{\left[\pdv{\alpha}f(\Gamma_v)  \right]\braket{\phi}}{\braket{\psi}}+F_{P}\left( \tilde{v}-v \right)
    \end{aligned}
\end{align}
with $v$ and $\tilde{v}$ defined as above.

In comparison to the minimal ansatz, the expression for the optimized ansatz contains Pfaffians of certain rows and columns of the resulting covariance matrix in  $\pdv{\alpha}f(M)$.
These are the results of Wick contractions from four-body correlators.
Since $f(M)$ is a sum of independent terms, we only treat a single Pfaffian explicitly
\begin{align}
    \begin{aligned}
        &\pdv{\alpha}\Pf\left(\Gamma_v(\alpha)\vert_{i_1,i_2,i_3,i_4}\right)\\
        &=\frac{1}{2}\Pf(\Gamma_v\vert_{i_1,i_2,i_3,i_4})\Tr \left( \Gamma_v\vert_{i_1,i_2,i_3,i_4}\inv \pdv{\Gamma_v\vert_{i_1,i_2,i_3,i_4}}{\alpha} \right)\, .
  \end{aligned}
\end{align}

For both, $F=1$ and $F=2$, the derivative of the electric energy contains the derivatives of the covariance matrix $\Gamma_v$.
The expression reads
\begin{align}
    \begin{aligned}
        \pdv{\alpha} \Gamma_v=&\pdv{\alpha}\left[ D_{oo}(\alpha)+D_{oc}(\alpha)(\tilde{D}(\alpha)-\gammainmod)\inv D_{oc}(\alpha)\tran \right]\\
        =&\pdv{D_{oo}}{\alpha}+\pdv{D_{oc}}{\alpha}  \left( \tilde{D}-\gammainmod \right)\inv D_{oc}\tran \\
        &-D_{oc}\left( \tilde{D}-\gammainmod \right)\inv\pdv{\tilde{D}}{\alpha} \left( \tilde{D}-\gammainmod \right)\inv D_{oc}\tran \\
        &+D_{oc}\left( \tilde{D}-\gammainmod \right)\inv\pdv{D_{oc}}{\alpha}\tran\, ,
    \end{aligned}
\end{align}
where we used $\pdv{K\inv}{\alpha}=-K\inv \pdv{K}{\alpha} K\inv$.

Similar to the idea of tracking $(D\inv-\gammain)\inv$, we can also track $(D-\gammain)\inv$ and avoid the expensive matrix inversions in each measurement computation.

The derivative of the matrices $D_{oo}$, $D_{oc}$,$D$ and $D$ are derivatives of the covariance matrix of the Majorana modes $D$.
As described in Appendix~\ref{app:covariance_matrices}, we know a direct construction of $D$ in terms of $T$. 
Thus, we can calculate the derivatives symbolically before the actual computation and insert the appropriate parameters as needed.
\end{document}